\documentclass[draftcls,12pt,onecolumn]{IEEEtran} 
\usepackage{booktabs}
\usepackage{multirow}
\usepackage{amsfonts}
\usepackage{amssymb}
\usepackage{amsbsy} 
\usepackage{amsmath}
\usepackage[ansinew]{inputenc}%
\usepackage{cite}
\usepackage{multirow}
\usepackage{color,xcolor}
\usepackage{lscape}
\usepackage[dvips]{graphicx}
\usepackage{colortbl}
\usepackage{url}
\usepackage{algorithm}
\usepackage{algorithmic}
\usepackage{stmaryrd}
\usepackage{hyperref}

\def\x{\ensuremath{\boldsymbol{x}}}

\def\R{\mathbb{R}}

\newcommand{\red}[1]{\textcolor[rgb]{1,0,0}{#1}}
\newcommand{\green}[1]{\textcolor[rgb]{0,0.5,0}{#1}}

\newcommand{\blue}[1]{\textcolor[rgb]{0.0,0.2,0.55}{#1}}
\newcommand{\mitch}[2]{\textcolor[rgb]{0.06,0.6,0.06}{#2}}

\newcommand{\mytitle}{Decision fusion with multiple spatial supports by conditional random fields}

\begin{document}

\title{\mytitle}
\author{Devis~Tuia,~\IEEEmembership{Senior Member,~IEEE}, Michele~Volpi,~\IEEEmembership{Member,~IEEE}, Gabriele Moser,~\IEEEmembership{Senior Member,~IEEE}
\thanks{\noindent DT was with the MultiModal Remote Sensing group, University of Zurich, Switzerland. He is now with the GeoInformation Science and Remote Sensing Laboratory, Wageningen University \& Research, the Netherlands. Email: devis.tuia@wur.nl.}
\thanks{\noindent MV is with the MultiModal Remote Sensing group, University of Zurich, Switzerland. Email: michele.volpi@geo.uzh.ch.}
\thanks{\noindent GM is with the Department of Electrical, Electronic, Telecommunication Engineering, and Naval Architecture, University of Genoa, Italy. Email: gabriele.moser@unige.it.}
\thanks{\noindent Digital Object Identifier  10.1109/TGRS.2018.2797316}}

\markboth{IEEE Transactions on Geoscience and Remote Sensing, Preprint, full version: 10.1109/TGRS.2018.2797316}{Tuia et al.: \mytitle}

\maketitle

\begin{abstract}
\noindent \textbf{This is the pre-acceptance version, to read the final version published in the IEEE Transactions on Geoscience and Remote Sensing, please go to: \href{https://doi.org/10.1109/TGRS.2018.2797316}{10.1109/TGRS.2018.2797316}}.
Classification of remotely sensed images into land cover or land use
is highly dependent on geographical information {at at least two
  levels}. {First,} land
cover classes are observed in a spatially smooth domain separated by
sharp region boundaries. {Second}, land classes and
observation scale are also tightly intertwined: they tend to be
consistent within areas of homogeneous appearance, or regions, in the
sense that all pixels within a roof should be classified as roof,
independently on the spatial support used for the
classification. \newline
In this paper we follow these two observations and encode them as
priors in an energy minimization framework based on conditional random
fields (CRFs), where classification results obtained at pixel and
region levels are probabilistically fused. The aim is to enforce the
final maps to be consistent not only in their own spatial supports
(pixel and region) but also \emph{across} supports, i.e. by getting
the predictions on the pixel lattice and on set of the regions to
agree. {Tho this end, we define an energy function with three
  terms: 1) a data term for the individual elements in each support
  {(support-specific nodes)}, 2) spatial regularization terms in
  a neighborhood for each of the supports {(support-specific
    edges)}, 3) a regularization term between individual pixels and
  the region containing each of them {(inter-supports edges)}.}
We utilize these priors in a unified energy minimization problem that
can be optimized by standard solvers. The proposed 2L$\lightning$CRF
model consists of a {CRF} defined over a bipartite graph, i.e. two
interconnected layers within a single graph accounting for
inter-lattice connections. 2L$\lightning$CRF is tested on two very
high resolution datasets involving submetric satellite and
subdecimeter aerial data. In all cases, 2L$\lightning$CRF improves the
result obtained by the independent base model (either random forests
or convolutional neural networks) and by standard CRF models enforcing
smoothness in the spatial domain.

\end{abstract}

\IEEEpeerreviewmaketitle


\section{Introduction}

Images with metric to decimetric spatial resolutions are becoming the new
standard for very high resolution (VHR) remote sensing: images
acquired by new generation satellites, as well as by sensors mounted
on aircrafts and drones, allow geometrically precise monitoring of
{the Earth surface} and can lead to breakthroughs in agriculture~\cite{Berni2009722}, forestry~\cite{Zahawi2015287}, urban characterization~\cite{DFCA}, and search-and-rescue~\cite{Tomic201246} tasks.

If the promise of very high resolution images is great, the tasks to
be solved also become harder: it is well known that as the spatial
resolution increases, so does the complexity of semantic labeling
(i.e. tasks aiming at assigning each pixel to a semantic
class). This is particularly true because of the increase of the
intra-class variance and {a parallel} decrease of the inter-class
variance. On one hand, a single semantic class is generally composed
of different materials (e.g. a building class is composed spectrally
of {several heterogeneous} materials found on the
roof). On the other hand different classes share the same materials
(e.g. vegetation is found in both forests and meadows). To address
such ambiguity, researchers in remote sensing have explored the use of
spatial context and structure, for which we will now review the main
families of solutions.

\paragraph{{Object-based image analysis}}
A first branch of {research} considered the
possibility to find a coarser spatial support for analysis {-- typically an agglomeration of pixels, a \emph{region} --} prior to learning a classifier. Also
known as \emph{Object-Based Image Analysis}
(OBIA~\cite{Blaschke2010}), this set of methods aim at defining a new
spatial support, possibly respecting the color gradients of the image,
extracting some textural and color features at the region level, and
learning a supervised classifier. {In computer vision, the task of finding coarser but
  semantically coherent spatial supports is often referred to as} \emph{superpixelization}~\cite{Fel04,Liu20112097,Achanta20122274},
where a set of coherent regions are found in the image, most often
without semantic meaning nor pretense to enclose a complete object in
a single region, but only assuming they contain parts of it
{and share parts of the borders}.

\paragraph{Spatial filters}
A second vivid research subfield is concerned with defining spatial
\emph{filters}, which are image processing techniques applied to the
image bands and aiming at {characterizing locally}
relevant properties of the semantic objects to be labeled. For
instance, to encode the property that nearby pixels tend to be of the
same class, local convolutions implementing low pass filters can be
used, while morphological filters~\cite{Mur10} and texture
filters~\cite{Pac08} are often used to encode more complex local
relationships. Recently, powerful descriptors from computer vision such
as Fisher vectors~\cite{Huang2016,Budak20161079},
bag-of-visual-words~\cite{Zhu2016747,Tu20161817,Rey17,Vol15} or local
binary patterns~\cite{Konstantinidis2016} have been proposed to
extract spatial information for remote sensing semantic labeling at
VHR. But a recurrent problem is to select the right bands to be
filtered, the family of filters and their parameters. These are
tasks that can prove to be very complex for a non-specialist and whose
failure can lead to very suboptimal models. Recent research
{addresses} this issue by learning the relevant filters
in an unconstrained space of filter candidates~\cite{Tui13c}, or to learn further
combinations of such filters, therefore reaching, through the
hierarchy of filters, higher level concepts closer to the semantic
classes~\cite{Tui15}. Authors in \cite{Volpi2015b} build very large feature
sets containing all the information which is possibly useful to
characterize the classification problem and then use a classifier that
is robust to unimportant features (i.e. embedding feature
selection). 

\paragraph{Convolutional neural networks}
The need to learn the relevant filters and features, instead of
pre-defining them by expert knowledge, focused recent research in
\emph{deep learning} as a tool for automatically learning multi-scale,
nonlinear, semantically tailored, and problem-specific contextual
filters jointly with a classifier. \emph{Convolutional neural networks}
(CNNs~\cite{lecun1998pieee}) are the perfect tool for solving such
task: they learn unconstrained convolutional filters
hierarchically. The filters learned in the first level (the first
\emph{convolutional layer}) of the network correspond mostly to local
oriented gradients, while those learned at the last convolutional
layers have a much wider receptive field (they {depend
  on} a wider area of the original images) and, therefore, inform
about a larger spatial context, usually semantically more related to
the problem to be solved. Since the filters are directly learned by
backpropagation, no effort has to be provided for filter engineering;
but all effort is moved to the design of the network structure itself
(number of layers, {type of} nonlinearities, spatial pooling,
size of filters, etc.). For a comprehensive introduction to CNN
building blocks in remote sensing, please see~\cite{Vol17}.
\newline
Thanks to numerous benchmark data released recently, the development
of CNN architectures for the semantic labeling of VHR aerial images
has flourished~\cite{Zhu17}: the first attempts (e.g.~\cite{DFCA})
performed inference in a sliding window fashion, therefore retrieving
the complete map pixel by pixel. This way one could use a standard
classification architecture to map from a patch to a single label
(supposed to represent the pixel on which the patch is centered on)
and retrieve the whole test map one patch at a time. Besides being
inefficient, this also limits the power of the CNN to encode spatial
information. A network learned for dense prediction (i.e. outputting
the entire map for each pixel in the patch) implicitly encodes spatial
relations between the different classes and their features at
different scales, while a network predicting a single label and
assigning it to the central pixel will consider each patch centered on
each pixel independently.
\newline Dense prediction by CNN, for instance inspired by fully
convolutional networks~\cite{long2015cvpr}, hypercolumns~\cite{Har15} or
SegNet~\cite{badrinarayanan2015segnet}, was proposed in {the
  field of computer} vision to upscale the CNN last layers to the
original input resolution and {\it de facto} provided an elegant,
fully learnable, solution to the problem of dense semantic
labeling. As a consequence, it was quickly adopted in remote sensing:
in~\cite{Vol17,Aud16,Marm16}, authors propose to use learned
deconvolution layers to upsample the activations, while
in~\cite{maggiori2017convolutional} the activations are simply
upsampled by interpolation. In~\cite{Mag16,Mar17b}, authors stack
upsampled activations at multiple scales to train other layers
performing dense prediction. Finally, in~\cite{sherrah2016arxiv}
authors propose a network without spatial poolings, i.e. a network
that preserves the spatial resolution of the original data throughout
all its components.
\newline But besides the impressive performances reported, most
approaches i) still showed some residual class inconsistencies and ii)
disregarded explicit object structure that could be easily integrated
by considering the superpixelization discussed above or some degree of
spatial reasoning (e.g. a building rarely occurs in the middle of a
river).

\paragraph{Structured prediction}
The last family of methods addresses this last point by encoding
spatial reasoning via interaction potentials between spatial units. 
Models such as Markov~\cite{GemanGeman,Bes92,JosianeZoltan} and
conditional random fields~\cite{Laf01,SuttonMcCallum} (MRFs or CRFs)
can be used to {jointly account for} different kinds of
prior information about spatial relationships and local
likelihoods. It has been used extensively in remote sensing to encode
{different kinds of assumptions one has about the data,
  such as} local spatial smoothness~\cite{Mos13,Sch12,LiMarkov} or, linearity~\cite{Smi97} or
3D arrangements~\cite{Luo2014}. 
Owing to the Hammersley-Clifford theorem~\cite{GemanGeman}, random
field models make it possible to express the posterior joint
distribution of the unknown class labels given the observations as a Gibbs
distribution. This leads to formulating the maximum {\it a-posteriori}
(MAP) criterion as the minimization of an appropriate energy function,
{modeling conditional relationships
  across variables.} The energy can be defined by a number of factors,
generally including local likelihood scores from a (set of)
classifier(s)~\cite{SolbergJain} and models of spatial relations
across locations and their labels. These relations usually encode the
likelihood of a pixel to belong to a class, given the class of its
direct neighbors~\cite{Vol15,Mar16}, or aim at capturing co-occurrence
structures among the classes~\cite{Hob15,Volpi2015b,Weg16}. In
addition, the definition of suitable Markovianity properties on
hierarchical graphs, including quad-trees, binary partition trees, or
more irregular topologies~\cite{Willsky,JosianePeppeRaffaele}, makes
it possible to also formalize multiscale and multiresolution fusion
within an MRF/CRF framework. Recent approaches based on random fields
have been proposed for the semantic segmentation of data acquired at
multiple input resolutions at the same time or in a time
series~\cite{IoMultires,Ihsen}. Finally, higher order random field
models exist~\cite{Koh08}, that can be used to encode higher order
relationships between elements forming a suitable group (named
clique): in remote sensing, such reasoning was exploited to detect
roads~\cite{JanHigherOrder} or relationships between different types
of classes occurring on different cliques, such as land cover and
land use~\cite{Albert201763}. Even if these models allow to encode
much more complex relationships, 
they also involve large computational load to perform inference.

{\paragraph{Co-segmentation} Our method also shares similarities
with the domain of image co-segmentation. Research in this direction is
often framed as a weakly supervised task, where labels for semantic
segmentation or object detection are not dense over the pixel lattice
but only appear as general "presence" or "absence" attributes for a given
class. It is also commonly phrased as an unsupervised step, where
objects belonging to the same object class in different images have to
be clustered together. Co-segmentation aims then at building models
able to extract common objects in a series of images under the prior
that the same objects are present \cite{rother2006cvpr,joulin2010cvpr}.
\newline
In this work, we build on the assumption that same classes are present
at the same locations, rather than at different locations with
possibly different viewpoints. We treat the class label for a given
spatial coordinate as the random variable to be fitted from a set of
semantic classes, and we do not additionally fit a binary segmentation
masks (foreground background) or bounding boxes while grouping the same
classes \cite{rother2006cvpr}. This simplification has also been used
in \cite{xiao2017tgrs}, where co-segmentation of the
multitemporal image pair is driven by the pixelwise difference image.}

\vspace{.3cm}

In this paper we aim at fusing the advantages of these families of
techniques by proposing a model that accommodates different spatial
supports {(or lattices, connectivity graphs)} and encodes
spatial reasoning relevant to the problem. We integrate the pixel- and
region-based strategies within a multi-scale approach, letting two (or
more) spatial supports {interact} and come to a
common decision. {The likelihood of each} granularity
level corresponds to a classification model on a given spatial
support. These responses at all levels are fused probabilistically
using a bipartite CRF (i.e. a CRF with two interconnected
layers~\cite{Hua08}).
\begin{figure}[!t]
\includegraphics[width=\linewidth]{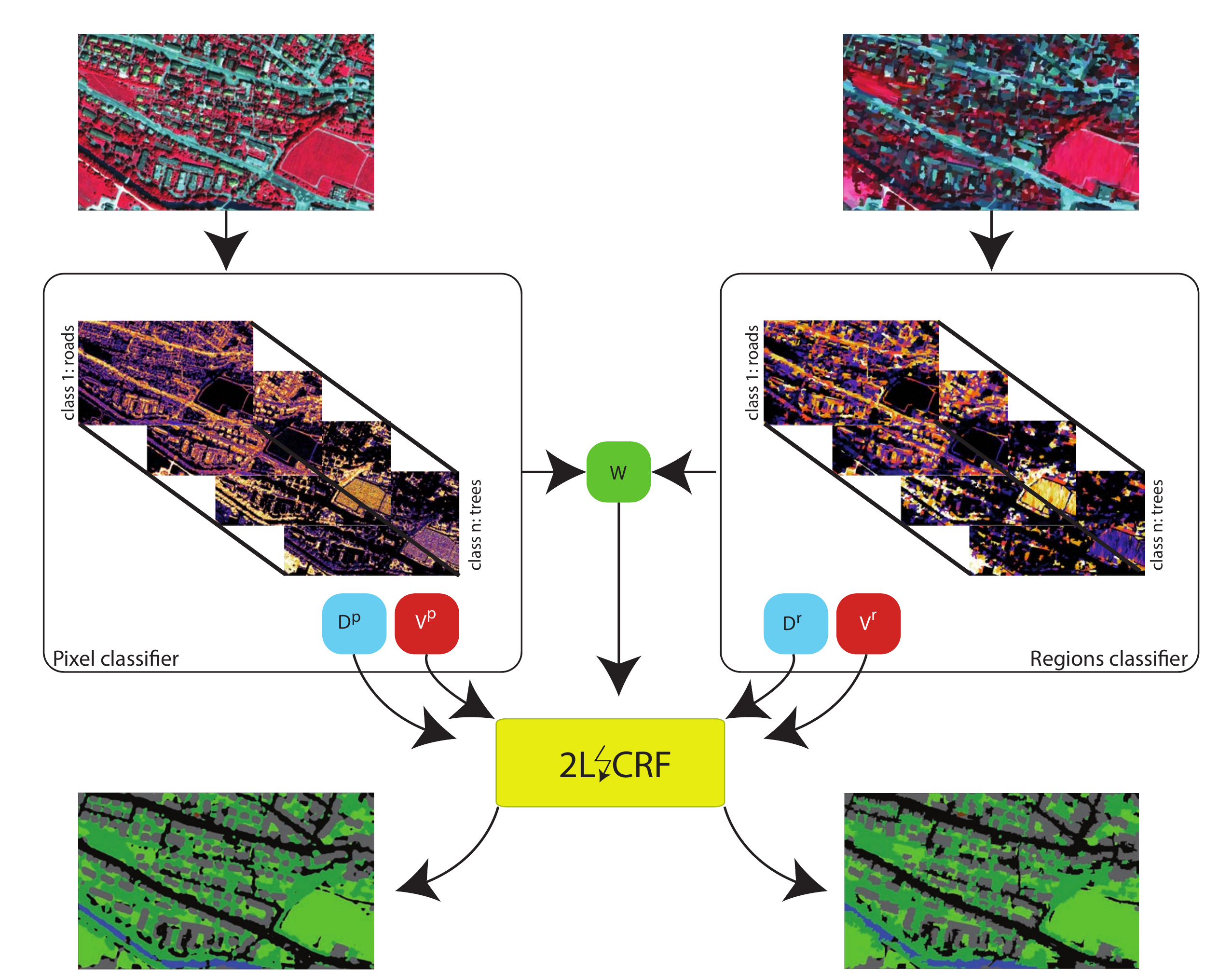}
\caption{The proposed 2L${\lightning}$CRF: starting from two models trained on different spatial supports, it uses the posterior probability scores $D$, spatial relations $V$ and hierarchical information $W$ to find the most likely maps (colors and symbols refer to Eq.~\eqref{eq:E}) }\label{fig:intuition}
\end{figure}
The model, named 2L${\lightning}$CRF {(read `two-layered-lightning-CRF'\footnote{{The lightning symbol ``$\lightning$'' is often used in proofs when contradictory statements are involved. Here, it is used on purpose to emphasize that the proposed CRF model is aimed at fusing the two layers, resolving contradictions between their prediction, and ``getting pixels and regions'' to agree on the final labeling.}})}
, is sketched in Fig.~\ref{fig:intuition}. 2L${\lightning}$CRF can use
as inputs the output of any classifier providing a distribution
over the labels: {they can be obtained
  by using} pre-defined spatial features {into standard
  classifiers} or {by} training CNNs. 2L$\lightning$CRF performs
structured prediction accounting simultaneously for a pixel- and a
region-based {granularity} by encoding 1)
spatial structuring via contrast-sensitive pairwise
potentials~\cite{Boykov2001} and 2) consistency between the labelings
at the pixel and region {lattices} via an
inter-layer smoothness assumption. Contrarily from higher order models
such as the P$^\text{n}$Potts~\cite{Koh08}, our conditional random
field avoids the multiscale structure of the graph {and the use
  of auxiliary variables} by encoding the multiscale problem as a
single (flat) graph with interscale connections, thus allowing to use
standard and computationally efficient energy minimization solvers. It
is more efficient that the iterative formulation of~\cite{Roig11} (which is the base
of~\cite{Albert201763}), since it solves both problems within a single
energy minimization step.
\newline A preliminary version of this study was previously introduced
by the authors in a conference paper~\cite{Tui16b}. We extend it to
the present paper, where we provide an in-depth methodological
analysis, a study of the impact of CNN unary scores to the system and
results on two datasets, Zurich Summer~\cite{Vol15} and the
challenging 2015 Data Fusion Contest dataset~\cite{DFCA}.

In Section~\ref{sec:hcrf} we present the formulation of the proposed
model, as well as the algorithm used to solve the energy minimization
problem. In Section~\ref{sec:data} we present the datasets and the
setup of experiments discussed in
Section~\ref{sec:res}. Section~\ref{sec:c} concludes the paper.


%
\newcommand{\I}{\mathcal{I}}
\newcommand{\X}{\mathcal{X}}
\newcommand{\Y}{\mathcal{Y}}
\newcommand{\G}{\mathcal{G}}
\newcommand{\V}{\mathcal{V}}
\newcommand{\E}{\mathcal{E}}
\newcommand\up[1]{{#1}^\uparrow}

\section{The proposed 2L$\lightning$CRF method}\label{sec:hcrf}
\subsection{Conditional random fields for semantic labeling}

In the framework of semantic labeling, conditional random fields
represent a family of probabilistic models {allowing to jointly
  characterize pixelwise class statistics as well as spatial
  dependencies between the labeling of different neighboring
  locations.} They are among the most used probabilistic graphical
models for remote sensing image analysis.

Let us consider a remote sensing image, from which $m$ features have
been extracted. The image presents $C$ thematic classes provided with
training samples. Let $\I$ be the regular pixel lattice, and $\x_i$
and $y_i$ be the feature vector and the class label of the $i$th pixel
($i\in\I;\,\x_i\in\R^m$;\,{$y_i {\in} \{1,2,\ldots,C\}$}). The
CRF approach considers $\x_i$ and $y_i$ as samples from two random
fields, i.e. a (generally continuous-valued) random field
$\X=\{\x_i\}_{i\in\I}$ of feature vectors and a discrete-valued random
field $\Y=\{y_i\}_{i\in\I}$ of class labels. Both random fields are
supported on the pixel lattice $\I$, on which a neighborhood system
$\{\partial i\}_{i\in\I}$ is
defined~\cite{JosianeZoltan}. Common choices include the first- and
second-order neighborhood systems, in which $\partial i$ is made of
the four pixels adjacent to the $i$th pixel and the eight pixels
surrounding it, respectively~\cite{LiMarkov}. With these notations,
the random field $\Y$ of the class labels is said to be a CRF if the
following posterior Markovianity property holds:
\begin{equation}\label{eq:CRFdef}
P(y_i|y_j,j\neq i,\X)=P(y_i|y_j,j\in\partial i,\X),
\end{equation}
and if the (global, imagewise) posterior distribution $P(\Y|\X)$ is
strictly positive~\cite{LiMarkov,SuttonMcCallum}. Condition
(\ref{eq:CRFdef}) means that the labels are spatially Markovian when
conditioned to the random field of the feature vectors. This property makes it possible
to express the posterior distribution as
$P(\Y|\X)\propto\exp[-U(\Y|\X)]$, where the energy $U(\Y|\X)$ is
defined according to the neighborhood
system~\cite{LiMarkov,GemanGeman}. {Minimizing such energy corresponds to the maximum a posteriori (MAP) criterion. For a system with only pairwise nonzero clique potentials, such energy can be written as~\cite{LiMarkov}: }
\begin{equation}\label{eq:CRF}
U(\Y|\X)=\sum_{i\in\I}D_i(y_i|\X)+\lambda\sum_{i\in\I}\sum_{j\in\partial i}V_{ij}(y_i,y_j|\X),
\end{equation}
where $D_i(y_i|\X)$, named \emph{unary} or association potential, is related
to the statistics of each individual label
given the feature random field. $V_{ij}(y_i,y_j|\X)$, named \emph{pairwise}
or interaction potential, encodes the spatial relations among the
labels of neighboring pixels ($i\in\I,j\in\partial i$). $\lambda$
is a positive parameter that tunes the tradeoff between the
unary and pairwise terms. The possibility to characterize the desired
spatial interactions by defining suitable pairwise potentials,
tailored to the application considered, and the availability of
computationally efficient energy minimization algorithms (see also
Section~\ref{sec:flattening}) make CRF modeling a powerful and
flexible approach to structured prediction in remote sensing image
analysis~\cite{Mos13,Sch12}.

\subsection{Overview and methodological assumptions of 2L$\lightning$CRF}

The key idea of the 2L$\lightning$CRF method is to benefit from both
pixelwise and region-based image representations by introducing a
novel model that connects two CRFs, defined on the pixel lattice and
on a segmentation result, to perform structured prediction at both
levels simultaneously. Given a VHR image, a segmentation method is
first applied to identify a set of homogeneous regions. An arbitrary
segmentation technique can be used within 2L$\lightning$CRF, provided
that the resulting segments are not particularly coarse.

We denote explicitly with a superscript ``$p$'' the quantities
introduced in the previous section at the granularity of individual
pixels ($\I^p,\partial^p i,\X^p,\Y^p$, etc.). Let $\I^r$ be the set of
regions resulting from segmentation, $\x^r_k$ be an $n$-dimensional
feature vector extracted from the image data of the $k$th region, and $y^r_k$ be the class
label of the same region ({$\x^r_k\in\R^n;y^r_k {\in} \{1,2,\ldots,C\};k\in\I^r$}). This
is equivalent to introducing a second pair of random fields
$\X^r=\{\x^r_k\}_{k\in\I^r}$ and $\Y^r=\{y^r_k\}_{k\in\I^r}$ that
collect feature vectors and labels at the granularity of regions. The
proposed method formalizes the relation between labels and feature
vectors at each granularity layer as a CRF, and merges these two CRFs
into a unique energy (see Section~\ref{sec:lightning}). 

From the probabilistic graphical modeling viewpoint, this means using
a bipartite graph to combine the two layers and generate a unique
labeling at the spatial resolution of the pixel lattice. In the
language of data fusion, the method fuses the information associated
with pixelwise statistics and with two sources of spatial information,
i.e. local neighborhoods and region-based reasoning. From a
computational perspective, the resulting energy is also equivalent to
a case-specific single-layer model on a planar graph -- a property
that makes it possible to use time-efficient algorithms to numerically
address the energy minimization task (see
Section~\ref{sec:flattening}).

More precisely, 2L$\lightning$CRF is based on the following
methodological assumptions:
\begin{enumerate}
\item In addition to the neighborhood system $\{\partial^p
  i\}_{i\in\I^p}$ on the pixel lattice, a neighborhood system
  $\{\partial^r j\}_{j\in\I^r}$ is also defined on the set of regions.
\item The random field $\Y^s$ of the labels at each granularity layer
  $s\in\{p,r\}$, given the corresponding random field $\X^s$ of the
  feature vectors, is a CRF with up to pairwise nonzero clique
  potentials.
\item The following conditional independence {assumption holds:
\begin{equation}\label{eq:condindpdf}
f(\X^p,\X^r|\Y^p,\Y^r)=f(\X^p|\Y^p)f(\X^r|\Y^r),
\end{equation}
where $f(\X^p,\X^r|\Y^p,\Y^r)$ is the joint probability density
function (PDF) of all feature vectors conditioned to all labels in
both granularity layers, and $f(\X^p|\Y^p)$ and $f(\X^r|\Y^r)$ are the
class-conditional PDFs associated with the two layers separately.}
\end{enumerate}

\begin{figure}[!t]
\includegraphics[width=.97\linewidth]{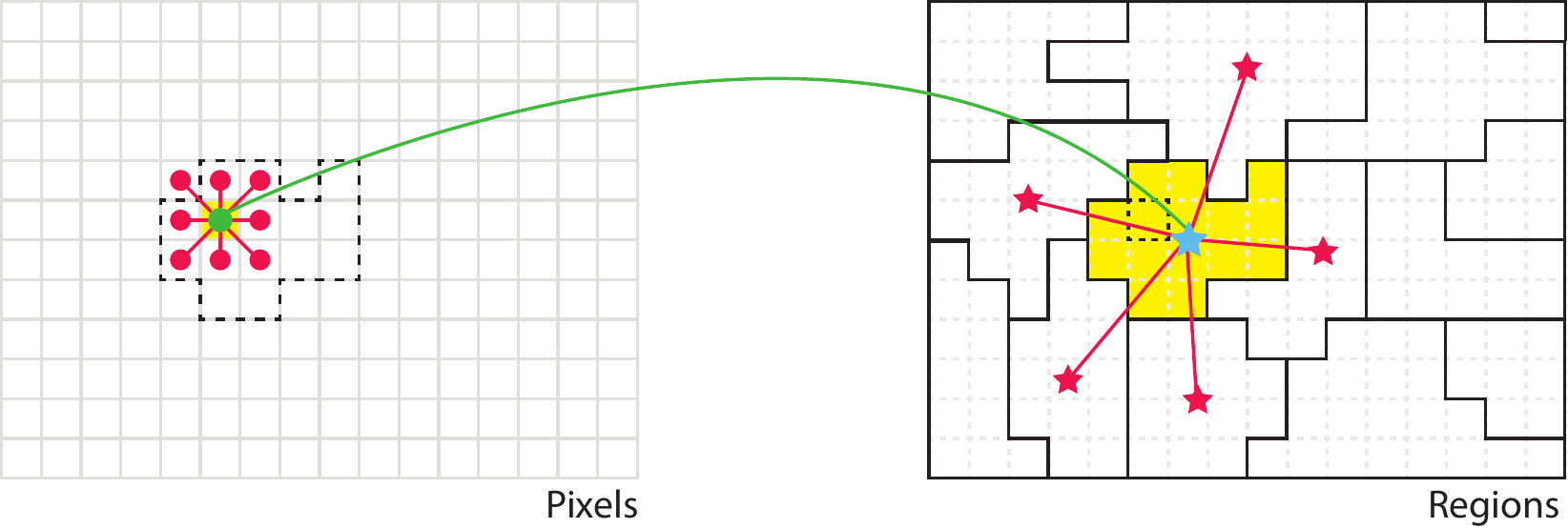}
\vspace*{-3mm}
\caption{General idea of the 2L$\lightning$CRF structured prediction model used to connect the pixel lattice and the set of regions. Red links represent the pairwise contrast-sensitive potentials (\textbf{[\red{B}]} terms in (\ref{eq:E}) between pixels (represented by circles) or region centroids (represented by stars). The green link represents the cross-layer potential relating a region and each pixel included in that region (\textbf{[\green{C}]} term in (\ref{eq:E})).  The locations of the considered pixel and region in the other layer are highlighted with dashed black lines. 
\label{fig:flow}}
\end{figure}

In particular, a second order neighborhood is used on the pixel
lattice, and two regions are considered to be neighbors (see
Assumption 1) if they share some common boundary (see
Fig.~\ref{fig:flow}). {Equation (\ref{eq:condindpdf}) indicates
  that the statistics of the features in the two layers are modeled as
  independent when conditioned to the labels of each layer (see
  Assumption 3). From a modeling viewpoint, this condition allows the
  class-conditional statistics within each layer to be characterized
  separately. This conditional independence assumption ensures
  mathematical tractability and is often accepted in MRF-based (and
  more generally Bayesian) approaches to spatial-contextual,
  multisource, or multitemporal
  classification~\cite{LiMarkov,JosianeZoltan,SolbergJain,melgani2003markov,Swain}.}

\subsection{The proposed two-layers model}\label{sec:lightning}

To apply the MAP decision rule over pixels and regions simultaneously,
we consider the joint posterior distribution of the combined random
field $\Y=(\Y^p,\Y^r)$ of pixel and region labels, given the combined
random field $\X=(\X^p,\X^r)$ of all feature vectors on both
layers. {Based on Assumptions 1-2-3, we prove that this joint
  posterior $P(\Y|\X)$ can be expressed as a function of i) two
  separate contributions associated with the marginal posteriors
  $P(\Y^p|\X^p)$ and $P(\Y^r|\X^r)$ of the two granularity layers,
  which are modeled as CRFs (see Assumption 2); and ii) a cross-layer
  term that is related to the dependence between $\Y^p$ and $\Y^r$ and
  provides a prior on the desired relations between region labels and
  pixel labels.}

{The Bayes theorem implies that:
\begin{eqnarray}
\nonumber P(\Y|\X)&=&P(\Y^p,\Y^r|\X^p,\X^r)\propto\\
&\propto& f(\X^p,\X^r|\Y^p,\Y^r)\cdot P(\Y^p,\Y^r),
\end{eqnarray}
where the proportionality factor depends on the feature vector fields $\X^p$ and $\X^r$, but not on the label fields $\Y^p$ and $\Y^r$, hence it does not influence MAP decisions. Owing to Assumption 3 and using again the Bayes theorem, we obtain:
\begin{eqnarray}\label{eq:mapgab}
\nonumber P(\Y|\X)&\propto&P(\Y^p,\Y^r)\prod_{s\in\{p,r\}}f(\X^s|\Y^s)\propto\\
\nonumber &\propto&P(\Y^p,\Y^r)\prod_{s\in\{p,r\}}\frac{P(\Y^s|\X^s)}{P(\Y^s)}=\\
&=&\frac{P(\Y^p,\Y^r)}{P(\Y^p)P(\Y^r)}\cdot\prod_{s\in\{p,r\}}P(\Y^s|\X^s),
\end{eqnarray}
where the proportionality factor depends again only on $\X^p$ and
$\X^r$. Equation (\ref{eq:mapgab}) provides the desired factorization
of the joint posterior.}

{Accordingly, the MAP rule is expressed as the minimization of
  the following energy function with respect to $\Y$:}
\begin{eqnarray}\label{eq:E}
\nonumber U(\Y|\X)&=&\sum_{s\in\{p,r\}}\sum_{i\in\I_s}\underbrace{D^s(y^s_i|\X^s)}_{\mathbf{[\blue{A}]}}+\\
\nonumber &+&\lambda\sum_{s\in\{p,r\}}\sum_{i\in\I_s}\sum_{j\in\partial^s i}\underbrace{V^s(y^s_i,y^s_j|\X^s)}_{\mathbf{[\red{B}]}}+\\
&+&\mu\underbrace{W(\Y^p,\Y^r)}_{\mathbf{[\green{C}]}},
\end{eqnarray}
where $\lambda$ and $\mu$ are positive weight parameters and the three
terms highlighted have the following meaning:
\begin{itemize}
\item[\textbf{[\blue{A}]}] This is the unary potential contributing to
  the CRF model for each single-layer posterior $P(\Y^s|\X^s)$ in
  (\ref{eq:mapgab}). As a customary choice in many MRF- and CRF-based
  methods~\cite{LiMarkov,Vol15,Hob15}, we compute it as ($i\in\I^s$):
  \begin{equation}\label{eq:unary}
    D^s(y^s_i|\X^s)=-\ln\hat{P}^s(y^s_i|\x^s_i),
  \end{equation}
  where $\hat{P}^s(y^s_i|\x^s_i)$ is an estimate of the element-wise
  posterior probability of a pixel or region. It can be computed by
  training, at layer $s$, a classifier that provides a probabilistic
  output (e.g., a parametric or non-parametric Bayesian
  classifier~\cite{Bishop}, a random forest~\cite{Breiman}, a
  CNN~\cite{lecun1998pieee}, or the postprocessing of the output of a
  support vector machine~\cite{Vapnik} by algorithms such
  as~\cite{LinPairwiseCoupling}).

  Note that this model for the unary potential in 2L$\lightning$CRF is
  intrinsically {homogeneous}, as reflected by the
  absence of the subscript ``$i$'' in (\ref{eq:E}) and
  (\ref{eq:unary}), as compared to (\ref{eq:CRF}) (i.e., $D^s(\cdot)$
  instead of $D_i^s(\cdot)$)~\cite{LiMarkov}. {This homogeneity
    in the unary potential is not considered a critical restriction
    because the region-based analysis incorporates local spatial
    adaptivity {\it per se}.}

\item[\textbf{[\red{B}]}] This energy contribution is a pairwise term
  that favors spatial smoothness at each granularity level (see the
  red line in Fig.~\ref{fig:flow}). It completes the CRF model for
  $P(\Y^s|\X^s)$ along with the aforementioned unary term
  \textbf{[\blue{A}]}. $V^s(y^s_i,y^s_j|\X^s)$ is defined as a
  contrast-sensitive potential that extends the classical Potts MRF
  model in order to penalize that different classes are predicted for
  neighboring pixels or regions with similar feature
  vectors~\cite{Boykov2001}. In 2L$\lightning$CRF, contrast
  sensitivity is modeled using a Gaussian kernel $K(\cdot)$
  ($i\in\I^s,j\in\partial^s i$):
  \begin{equation}\label{eq:pairwise}
    V^s(y^s_i,y^s_j|\X^s)=[1 - \delta(y^s_i,y^s_j)]K(\x^s_i,\x^s_j),
  \end{equation}
  where $\delta(\cdot)$ is the Kronecker symbol (i.e., $\delta(a,b)=1$
  for $a=b$ and $\delta(a,b)=0$ otherwise). For instance, if two
  neighboring pixels have identical feature vectors (hence
  $K(\x^p_i,\x^p_j) = 1$) and are predicted in different classes, then
  the maximum penalty is applied. On the contrary, if their feature
  vectors differ substantially (thus $K(\x^p_i,\x^p_j)\simeq 0$), then
  assigning the two pixels to different classes is not (or very
  slightly) penalized. The same comment holds in the case of
  regions. The previous remark on {homogeneity}
  holds in this case (\ref{eq:pairwise}) as well.  \newline Note that
  the same parameter $\lambda$ is used in (\ref{eq:E}) to weigh both
  $V^p(\cdot)$ and $V^r(\cdot)$. This is consistent with the idea of
  giving both granularity layers the same relevance in the labeling
  process. Nevertheless, extending (\ref{eq:E}) with different weight
  parameters for the pixel and region granularities is
  straightforward.

\item[\textbf{[\green{C}]}] This energy term is a cross-layer pairwise
  contribution that favors consistency between the labelings at the
  two granularity levels (see the green line in Fig.~\ref{fig:flow})
  {and corresponds, up to additive or positive multiplicative
    constants, to $-\ln\{P(\Y^p,\Y^r)/[P(\Y^p)P(\Y^r)]\}$. This term
    is related to the joint prior $P(\Y^p,\Y^r)$ and provides a
    measure of the dependence between the fields $\Y^p$ and
    $\Y^r$. The rationale of [\green{C}]} is to encode the desired
  agreement between pixelwise and region-based results. This behavior
  is favored in the proposed method by using a Potts-like formulation:
  \begin{equation}\label{eq:cross}
    W(\Y^p,\Y^r)=\sum_{i\in\I^p}[1 - \delta(y^p_i,y^r_{\up{i}})],
  \end{equation}
  where $\up{i}$ indicates the region to which the $i$th pixel
  belongs, i.e. $y^p_i$ and $y_{\up{i}}^r$ are the labels of this
  pixel at the pixel and region levels, respectively ($i\in\I^p$).
  Indeed, \eqref{eq:cross} contributes a penalty for each pixel for
  which the classes predicted at the two levels differ.
\end{itemize}

{More generally, formulating the decision fusion over multiple
  supports as the minimization of energy (\ref{eq:E}) can also be
  interpreted as a variational representation, especially in relation
  to the belief-propagation-type method that is used to address it
  (see
  Section~\ref{sec:flattening})~\cite{SuttonMcCallum,Wainwright}.}

\begin{figure}[!t]
\includegraphics[width=.97\linewidth]{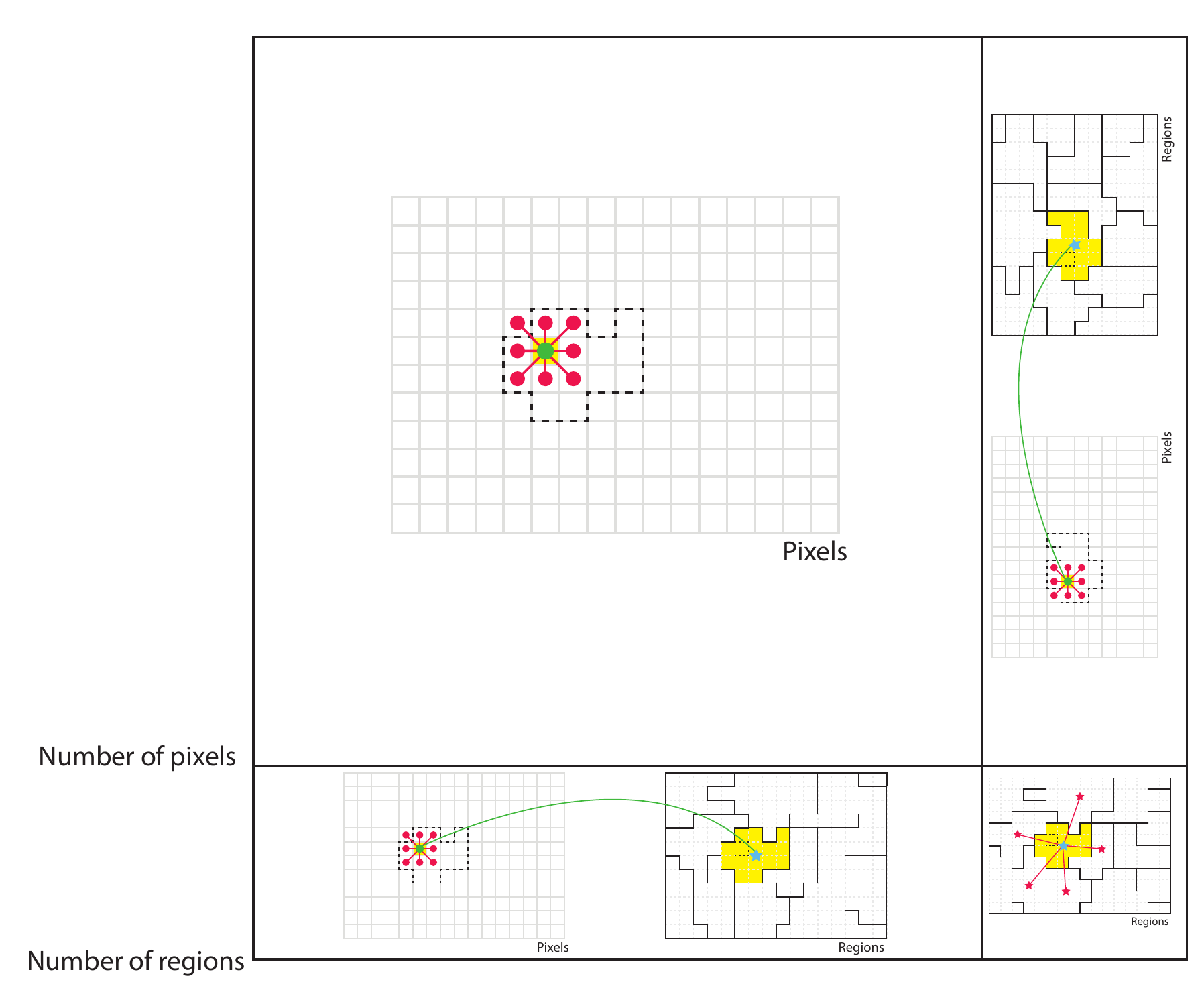}
\vspace*{-3mm}
\caption{Intuition behind the graph structure used in the dual layer.\label{fig:graph}}
\end{figure}

\subsection{Flattening the two layers into a single graph and minimizing the energy}\label{sec:flattening}

The problem of the minimization of CRF energy functions such as
(\ref{eq:CRF}) is generally a complex combinatorial
problem. Nevertheless, in addition to consolidated stochastic
optimization approaches such as simulated annealing~\cite{GemanGeman},
computationally efficient graph-theoretic algorithms have become very
prominent during the past
decade~\cite{szeliski2006comparative,LiMarkov}. They include graph cut
algorithms, which make use of a min-flow/max-cut reformulation of the
minimum energy problem~\cite{Boykov2001}, and belief propagation-type
algorithms, which build on the idea of exchanging messages between
neighboring elements to progressively reduce the
energy~\cite{kolmogorov2006convergent}. These techniques have been
found successful to minimize energies that are supported on a pixel
lattice (as in (\ref{eq:CRF})) or even on more general
graphs~\cite{szeliski2006comparative}. On one hand, in the case of the
proposed method, two layers with distinct supports are involved in the
minimization task simultaneously, a situation that does not allow
applying these methods directly. On the other hand, we show here that
2L$\lightning$CRF can also be equivalently reformulated on a unique
planar graph, thus making it possible to use the aforementioned energy
minimization methods appropriately.

The idea is to combine the two granularity layers into a single
undirected graph $\G=(\V,\E)$, in which both the spatial
[\textbf{\red{B}}] and the cross-layer [\textbf{\green{C}}] pairwise
interactions are reorganized. As shown in Fig.~\ref{fig:graph}, the
set of nodes is {$\V=\I^p\cup\I^r$} (i.e. it
includes all pixels and regions), and an edge exists between two nodes
$u$ and $v$, i.e., $(u,v)\in\E$, if and only if one of the following
conditions holds ($u,v\in\V$):
\begin{itemize}
\item $u$ and $v$ are neighboring pixels, i.e. $u\in\I^p,v\in\partial^p u$;
\item $u$ and $v$ are neighboring regions, i.e. $u\in\I^r,v\in\partial^r u$;
\item $u$ is a pixel and $v$ is the region that includes $u$,
  i.e. $u\in\I^p$ and $v=\up{u}$, or vice versa.
\end{itemize}
The random fields $\X^p,\X^r,\Y^p$, and $\Y^r$ on the pixel lattice
and on the set of regions can be rearranged on the graph $\G$ in a
straightforward way (see Fig.~\ref{fig:graph}). Denoting with a bar
the rearranged quantities (e.g., $\bar\Y=\{\bar{y}_v\}_{v\in\V}$,
where $\bar{y}_v$ indicates the label of node $v$ and is either
$y^p_v$ or $y^r_v$ as $v\in\I^p$ or $v\in\I^r$, respectively), the
energy (\ref{eq:E}) can be rewritten as:
\begin{equation}
  \bar{U}(\bar\Y|\bar\X)=\sum_{v\in\V}\bar{D}_v(\bar{y}_v|\bar\X)+
  \sum_{(u,v)\in\E}\bar{V}_{uv}(\bar{y}_u,\bar{y}_v|\bar\X),
\end{equation}
where the unary potential is:
\begin{equation}
\bar{D}_v(\bar{y}_v|\bar\X)=\begin{cases}
D^p(y_v^p|\X^p) & \text{if }v\in\I^p\\
D^r(y_v^r|\X^r) & \text{if }v\in\I^r
\end{cases}
\end{equation}
and the pairwise potential is:
\begin{eqnarray}
&&\bar{V}_{uv}(\bar{y}_u,\bar{y}_v|\bar\X)=\\
\nonumber&&=\begin{cases}
\lambda[1-\delta(\bar{y}_u,\bar{y}_v)]K(\bar\x_u,\bar\x_v) & \text{if }(u\in\I^p\text{ and }v\in\partial^p u)\\
&\text{or }(u\in\I^r\text{ and }v\in\partial^r u)\\
\mu[1 - \delta(\bar{y}_u,\bar{y}_v)] & \text{if }\left(u\in\I^p\text{ and }v=\up{u}\right)\\
&\text{or }\left(v\in\I^p\text{ and }u=\up{v}\right).\\
\end{cases}
\end{eqnarray}
{Note that, although the unary and pairwise potentials in
  (\ref{eq:unary}) and (\ref{eq:pairwise}) are homogeneous, the
  potentials $\bar{D}_v(\cdot)$ and $\bar{V}_{uv}(\cdot)$ on the
  flattened planar graph $\G$ are only piecewise homogeneous.}

To minimize the resulting energy $\bar{U}(\bar\Y|\bar\X)$ with respect
to $\bar\Y$ on the graph $\G$ (i.e. with respect to $\Y^p$ and $\Y^r$
simultaneously), the sequential tree re-weighted message passing
(TRW-S) algorithm is used in 2L$\lightning$CRF. This algorithm
integrates the belief propagation approach and the construction of
appropriate spanning trees on the considered graph, and makes use of a
specific sequential formulation to favor a convergent behavior
(details can be found in~\cite{kolmogorov2006convergent}).


%


\section{Data and setup}\label{sec:data}

{In this Section, we introduce and discuss the datasets
  used for the experiments. We also detail
  the setup of said experiments.}

\subsection{Data}

The performance of the proposed 2L$\lightning$CRF is tested {using} two very-high
resolution datasets: the Zurich Summer and the Zeebruges data{sets}.

\begin{figure}[!t]
\setlength{\tabcolsep}{1pt}
\includegraphics[width=.90\linewidth]{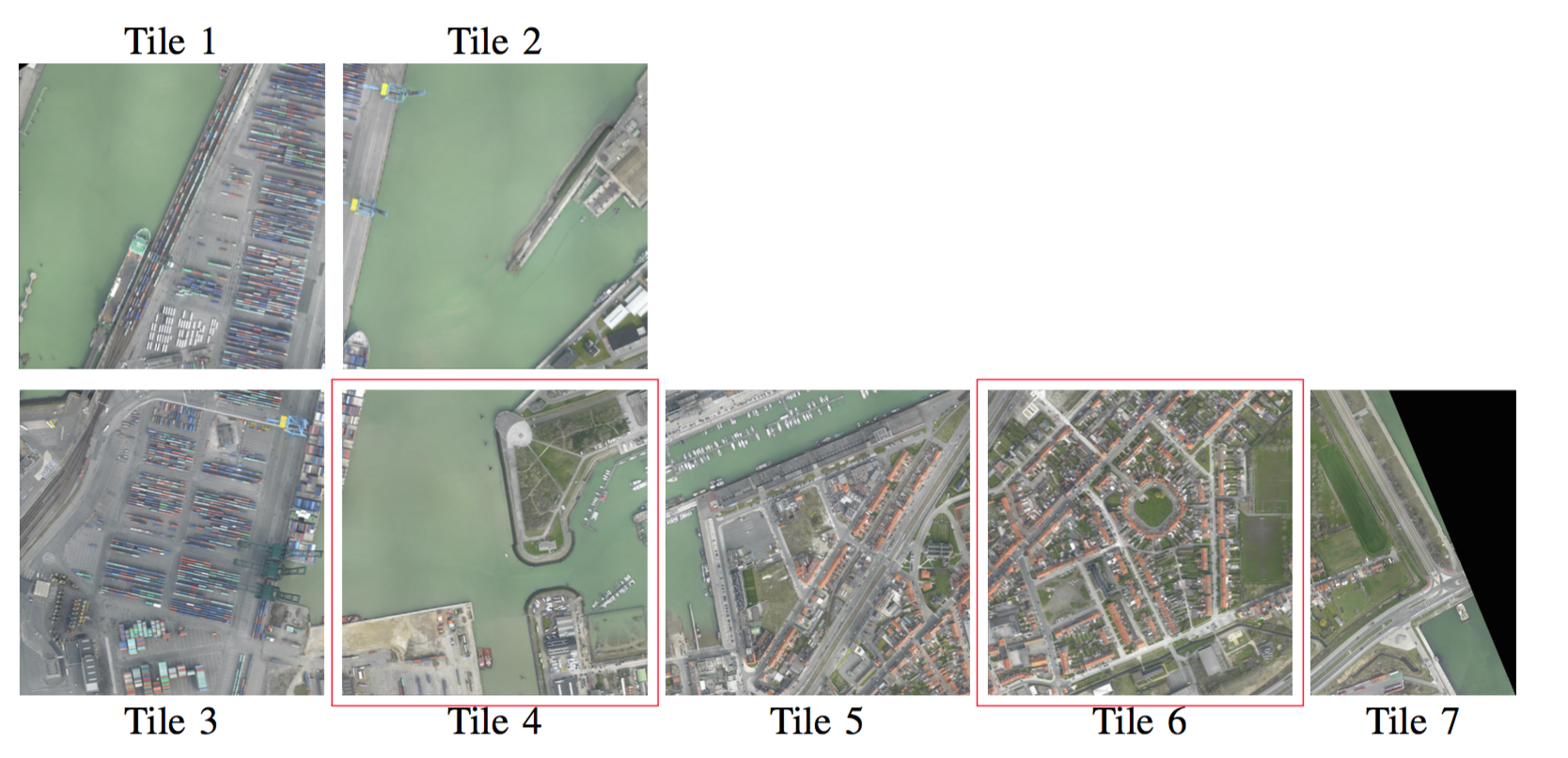}
\caption{The seven tiles of the Data Fusion Contest 2015 dataset (RGB) (from~\cite{DFCA}). Test tiles are highlighted in red. The data can be freely downloaded {from} \url{http://www.grss-ieee.org/community/technical-committees/data-fusion/2015-ieee-grss-data-fusion-contest/}. \textbf{(Note: in this preprint we had to decrease the graphics resolution. For full resolution, please refer to the published version, or contact the authors)}.}
\label{fig:tiles}
\end{figure}

\begin{figure}[!t]
\setlength{\tabcolsep}{1pt}
\includegraphics[width=.90\linewidth]{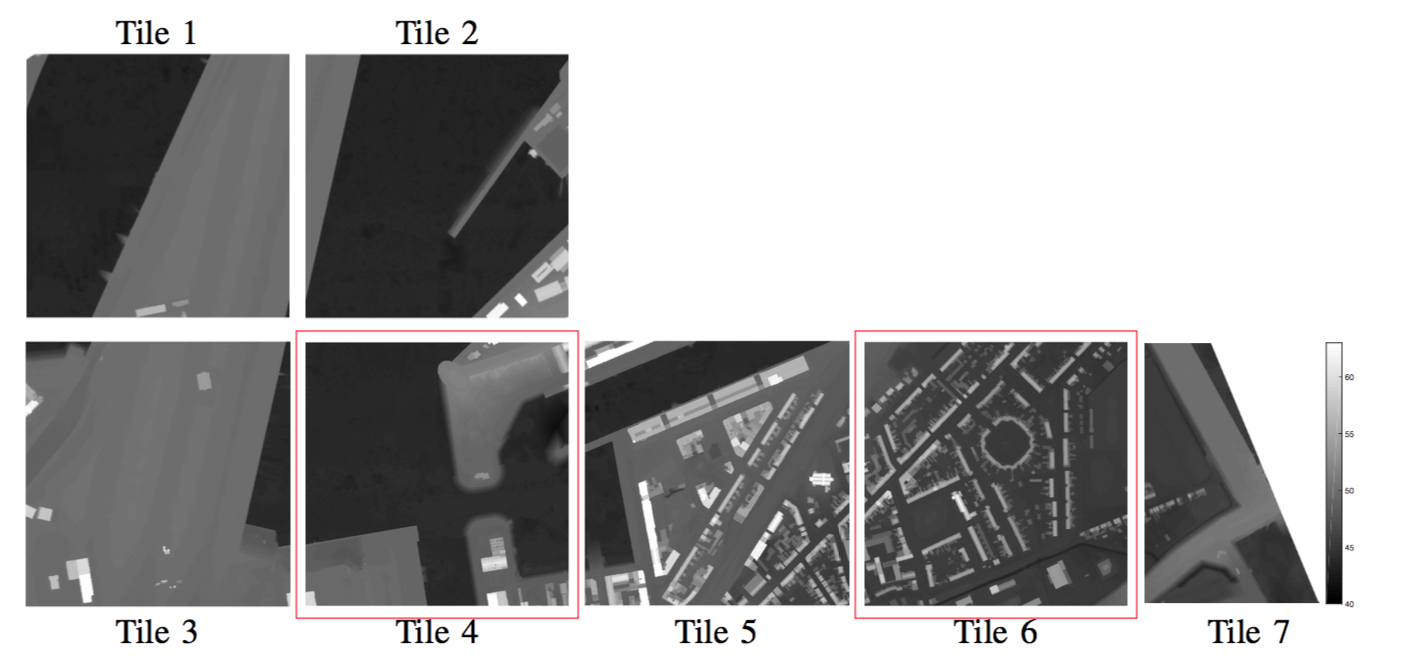}
\caption{The seven tiles of the DSM issued from the LiDAR point cloud of the Data Fusion Contest 2015 (from~\cite{DFCA}). Test tiles are highlighted in red. The data can be freely downloaded {from} \url{http://www.grss-ieee.org/community/technical-committees/data-fusion/2015-ieee-grss-data-fusion-contest/}. \textbf{(Note: in this preprint we had to decrease the graphics resolution. For full resolution, please refer to the published version, or contact the authors)}.}
\label{fig:tilesL}
\end{figure}

\begin{itemize}
\item[-] \emph{The Zurich Summer Dataset}~\cite{Volpi2015b}. This
  dataset is a collection of 20 images from a single large QuickBird
  (NIR-RGB bands) acquisition of 2002. The images picture different
  neighborhoods of the city of Zurich, Switzerland. Each image is
  pansharpened to 0.6~m resolution and labeled in 8 classes. The
  labeling is not dense, meaning that some pixels are either
  unassigned or belonging to unseen classes. False color infrared
  images and the ground truth for five tiles can be seen in the first
  two columns of Fig.~\ref{fig:resMaps}. The dataset can be freely
  downloaded at \url{https://sites.google.com/site/michelevolpiresearch/data/zurich-dataset}.

\item[-] \emph{Zeebruges, or the Data Fusion Contest 2015 dataset
    (\texttt{grss\_dfc\_2015})}~\cite{DFCA}. The Image Analysis and
  Data Fusion Technical Committee of the IEEE Geoscience and Remote
  Sensing Society organized in 2015 a data processing competition
  which aimed at 5-centimeter resolution land cover mapping. Both an
  RGB aerial image (5~cm spatial resolution) and a dense lidar point
  cloud (65~pts/m$^2$) were acquired over the harbor of Zeebruges,
  Belgium. The data are organized on seven {$10,000 \times 10,000$}
  pixels tiles. We used a downgraded version on {$5,000 \times 5,000$}
  pixels for training the models and upsampled the final prediction
  using interpolation (the final numerical scores were not
  significantly affected, but computational efficiency was
  dramatically improved). All the tiles have been densely annotated in
  8 land classes, including land use (buildings, roads) and small
  objects (vehicle{s}, boats) classes by the authors
  of~\cite{Lag15}. The seven RGB tiles, as well as the LiDAR
  normalized DSM are illustrated in Figs.~\ref{fig:tiles}
  and~\ref{fig:tilesL}. The ground truth test tiles are presented only
  in a blurred manner, since they are undisclosed to the participants
  of the challenge. The data can be obtained from the IEEE GRSS Data
  and Algorithm Standard Evaluation Website (DASE)
  \url{http://dase.ticinumaerospace.com/}. From DASE, users can
  download the seven tiles and labels for five tiles. To assess models
  on the two remaining tiles, we uploaded the classification maps on
  the DASE server, which automatically computes confusion matrices and
  accuracy scores. For more information about the data, please refer
  to~\cite{DFCA} and \url{https://www.grss-ieee.org/community/technical-committees/data-fusion/2015-ieee-grss-data-fusion-contest/}.
\end{itemize}

\subsection{Experimental setup}

The experiments presented in the next section aim at showcasing the
effectiveness of the proposed 2L$\lightning$CRF in different settings
using different classifiers. In all experiments, we compare the
results obtained using models aware of a single support (i.e. only
pixel- or region-based mapping units) with those obtained by models
exploiting structure in the spatial domain of the support (in all
cases, the CRF of~\cite{Koh08}) and with the results of the proposed
model, which lets the two supports interact via the two-layers
CRF structure. Below we present the setups used for the two datasets.

\subsubsection{Zurich summer}

We consider two base classifiers: random forests{~\cite{Breiman}}
and convolutional neural networks{~\cite{lecun1998pieee}} with
independent patch-response.

\begin{itemize}
\item[-] \emph{Random forest {(RF)}}. In the case of random forests,
  we train two separate models, one at the pixel and another at the
  region level~\cite{Tui16b}. At the pixel level, we use spectral
  features (R-G-B-NIR raw values, normalized to standard scores), {the
    normalized difference vegetation and water indices (NDVI and
    NDWI),} and contextual features (average over $3 \times 3$ and
  $5 \times 5$ local windows, computed on the R and NIR
  bands). Therefore, we have a 10-dimensional input space. {Regarding}
  the region level, we first extract regions using the Felzenszwalb
  and Huttenlocher graph-based algorithm~\cite{Fel04}. We used as
  region descriptors the min, max, average and standard deviation
  values of the pixel intensities included in each region for the R,
  G, B, NIR, NDVI, and NDWI channels. The input space of the region
  classifier is therefore 24-dimensional. To train the models, we
  derived a region ground truth from the available pixelwise training
  map by assigning the majority class found within each region.
\item[-] \emph{Convolutional neural network}. In the case of CNNs, we
  opted for a classical image classification architecture, i.e. an
  architecture that predicts a single label per patch {(in our
    case, of size $65 \times 65$ pixels)} and predicts the final map
  using a sliding window approach, as in~\cite{DFCA}. The reasoning
  behind it is that, since the ground truth is not dense, spatial
  structures become less explicit to be learned, and
  complex optimization schemes (as the one in~\cite{Vol15}), which are
  not trivial to stack on top of a dense prediction CNN, should be
  involved to learn them properly. Specifically, we use four
  convolutional blocks, the first two with $5 \times 5$ filters
  (expressed as two consecutive $3 \times 3$ filters, as proposed
  in~\cite{simonyan2015iclr}) and the other two with plain $3\times 3$
  filters. Each filter is followed by a ReLU nonlinearity and by a
  spatial pooling halving the spatial extent of the activations. The
  fourth block is followed by a fully connected layer outputting a
  $256$-dimensional vector per patch. This vector is then used to
  predict the class conditional probabilities with a soft-max
  classifier. A dropout~\cite{srivastava2014jmlr} rate of 50\% is used
  on the fully connected layer to reduce the risk of overfitting the
  training data. Given the computational efforts that would have been
  involved, we did not train a specific region-based model, but rather
  averaged and re-normalized the class probabilities per region and
  used them as scores for the region. {The results were not
    affected by using the averaged CNN pixelwise class probabilities
    instead of training a specific model for the region lattice (be it
    a CNN or the RF model presented above).}
\end{itemize}

In both cases, {the neighborhood systems used are a second
  order (8 direct neighbors) at the pixel level and the set of regions
  sharing a common boundary with the region under consideration at the
  region level.\newline W}e use images ID 1-15 (out of the 20 images
composing the dataset) to train the models and images 16-20 to test its
generalization. Accuracy figures are provided per image and averaged
over the test set, for both the pixel and region levels. For the case
of random forest we trained the pixel-level classifier with 0.01\% of
the available labeled pixels, with a selection stratified by class
(corresponding to $122,658$ pixels pooled from all the training
images). At the region level, the second RF classifier used all the
available $34,020$ labeled regions. The increase of training set size
at pixel level with respect to~\cite{Tui16b} is due to the intention
of providing a better numerical comparison against the CNN model, for
which {$10,000$} patches are used for training, randomly selected, and
re-sampled every $10$ training epochs. Since we run 150 epochs of the
model, the CNN actually saw roughly $150,000$ patches during the whole
training, whose diversity was also increased by data augmentation,
i.e. random flips and rotations at each 10 epochs resampling. The CNN
was trained by stochastic gradient descent with momentum (0.9), with a
weight decay of $0.001$ and a constant learning rate of {$10^{-3}$}.

\subsubsection{Zeebruges}

In the case of Zeebruges, we opted for the CNN case only, as Random
Forest do not perform nearly as accurate. In~\cite{DFCA}, two main
observations were made: CNN outperformed other nonlinear methods (such
as support vector machines) and a fully-trained model provided the
most accurate solution on this dataset. We also opted for a dense
prediction architecture: for this dataset, we have a dense ground
truth~\cite{Lag15} and therefore spatial relationships can be
explicitly learned. Following these two reasons, we used the CNN
in~\cite{Mar17b}, whose architecture follows the concept of
hypercolumns~\cite{Har15}, but with added equivariance to rotation.
The CNN is based on a set of convolutional blocks with spatial pooling
and all the activations at each level are upsampled to the original
image resolution, stacked and then fed into a second block employing
this multiscale descriptions to map to class likelihoods.
\mitch{used to train a traditional Multi-layer perceptron
  (MLP) providing the pixel scores.}{} In our case, the convolutional
blocks are RotEqNet layers enforcing rotation equivariance of the
final prediction (see~\cite{Mar17} for details). There are six such
blocks, learning respectively $[14,14,21,28,28,28]$ filters of size $7
\times 7$ each. After concatenation of the activations, the CNN has
three $1 \times 1$ convolution layers with $350$, $350$, and $6$ (the
number of classes) neurons, and a softmax normalization. The model was
trained for 34 iterations with decreasing weight decay (from
{$10^{-1}$} to $4 \cdot10^{-3}$) and learning rates (from {$10^{-2}$}
to $4\cdot 10^{-4}$).


Finally, for both datasets hyperparameters $\mu$ and $\lambda$ were
set by crossvalidation and the $\sigma$ parameter of the kernel in the
contrast sensitive term (Eq.~\ref{eq:pairwise}) was set as half the
Euclidean features distance between samples.


%


\definecolor{resi}{RGB}{102,102,102}
\definecolor{street}{RGB}{0,0,0}
\definecolor{tr}{RGB}{51,155,0}
\definecolor{mea}{RGB}{0,255,0}
\definecolor{rail}{RGB}{255,255,0}
\definecolor{wat}{RGB}{0,0,255}
\definecolor{pool}{RGB}{102,153,255}
\definecolor{bare}{RGB}{128,64,0}

\definecolor{Zimp}{RGB}{1,1,1}
\definecolor{Zwater}{RGB}{0,0,125}
\definecolor{Zclutter}{RGB}{255,0,0}
\definecolor{ZlowVeg}{RGB}{0,255,255}
\definecolor{Zbuilding}{RGB}{0,0,255}
\definecolor{Ztree}{RGB}{0,255,0}
\definecolor{Zboat}{RGB}{255,0,255}
\definecolor{Zcar}{RGB}{255,255,0}

\section{Experimental results}\label{sec:res}

In this section we present the results obtained on the Zurich Summer
and Zeebruges datasets.

\subsection{Zurich Summer dataset}\label{sec:resZH}
For the Zurich Summer dataset, we first discuss results obtained by the
approach considering random forests with contextual features and then
compare it against the results obtained by the CNN base classifier.

\noindent \textbf{Random forests.} Table~\ref{tab:numRF} shows the
numeric scores obtained using random forest. Results are
in line with those reported in~\cite{Tui16b}. The slight increases
in accuracies are due to the larger training set used to train the
pixel-based classifier and are consistent across the experiments. When
using random forest, the 2L$\lightning$CRF approach allows to increase
{consistently} the performances of the base
independent models with average increases of accuracy of around
4\%. An exception is found when considering the mean of the average
accuracy (AA) over the five tiles (`Ov.' in the table): in this case,
the 1.8\% decrease in AA observed is due to poor performance in tile
\#16 (- 5\%). The poor performance is explained by the disappearance
of the class `bare soil' from the prediction, which decreases
strongly the average accuracy. When pooling confusion matrices
over the five tiles (`Avg.' in the table) this effect disappears, as
bare soil is correctly predicted in the other tiles.


\newcommand{\mysize}{2pt}

\begin{table*}[!t]
\caption{Zurich Summer dataset: numerical results on the five test tiles. Each pair of columns illustrates the change of performance between the unaries ({r}andom forest) and the {spatial} model using them as a base {probabilistic input in the unary potential}. The CRF is based on~\cite{Boykov2001}.
\label{tab:numRF}}
\setlength{\tabcolsep}{\mysize}
\scriptsize{
\vspace{.1cm}
\begin{tabular}{c|ccl |ccl ||ccl |ccl ||ccl |ccl}
\cline{2-19}
      & \multicolumn{6}{c||}{Ov. Accuracy (OA)} & \multicolumn{6}{c||}{Kappa ($\kappa$)} &  \multicolumn{6}{c}{Av. Accuracy (AA)}\\\hline
Tile  & \multicolumn{3}{c|}{Pixels} & \multicolumn{3}{c||}{Regions} & \multicolumn{3}{c|}{Pixels} & \multicolumn{3}{c||}{Regions} & \multicolumn{3}{c|}{Pixels} & \multicolumn{3}{c}{Regions} \\
\#    & {RF} & {CRF}& {2L$\lightning$CRF} & {RF} & {CRF}& {2L$\lightning$CRF} & {RF} & {CRF}& {2L$\lightning$CRF} & {RF} & {CRF}& {2L$\lightning$CRF} &  {RF} & {CRF}& {2L$\lightning$CRF} &  {RF}& {CRF} & {2L$\lightning$CRF} \\
\hline
16	&	80.3	&	81.8	&	85.1	&	82.6	&	84.8	&	87.5	&	0.72	&	0.75	&	0.79	&	0.76	&	0.79	&	0.82	&	60.0	&	60.0	&	58.4	&	63.7	&	63.3	&	58.9	\\
17	&	77.7	&	79.2	&	83.4	&	82.9	&	83.3	&	86.3	&	0.71	&	0.73	&	0.78	&	0.77	&	0.78	&	0.82	&	59.8	&	60.6	&	62.8	&	69.3	&	68.3	&	64.5	\\
18	&	69.6	&	71.8	&	78.1	&	71.2	&	73.4	&	79.9	&	0.57	&	0.6	&	0.68	&	0.58	&	0.61	&	0.70	&	62.2	&	63.4	&	66.9	&	61.2	&	62.5	&	65.9	\\
19	&	67.4	&	68.6	&	69.5	&	65.5	&	66.8	&	69.4	&	0.57	&	0.59	&	0.60	&	0.55	&	0.57	&	0.60	&	60.0	&	60.5	&	60.5	&	68.7	&	70.3	&	62.2	\\
20	&	76.2	&	77.7	&	82.9	&	78.0	&	79.2	&	84.4	&	0.69	&	0.71	&	0.78	&	0.72	&	0.73	&	0.80	&	65.6	&	67.0	&	71.1	&	67.2	&	68.3	&	72.3	\\\hline
Avg.$^\dagger$	&	75.4	&	76.9	&	80.9 \green{(+4.0)}	&	77.3	&	78.8	&	82.7 \green{(+3.9)}	&	0.69	&	0.71	&	0.76 \green{(+0.05)}	&	0.71	&	0.73	&	0.78 \green{(+0.05)}	&	66.0	&	67.0	&	69.5 \green{(+2.5)}	&	67.1	&	67.8	&	70.5 \green{(+2.7)}	\\
Ov.$^*$	&	74.2	&	75.8	&	79.8 \green{(+4.0)}	&	76.0	&	77.5	&	81.5 \green{(+4.0)}	&	0.65	&	0.67	&	0.73 \green{(+0.06)}	&	0.68	&	0.70	&	0.75 \green{(+0.05)}	&	61.5	&	62.3	&	63.9 \green{(+1.6)}	&	66.0	&	66.6	&	64.8 \red{(--1.8)}	\\
\hline
\multicolumn{9}{l}{$^\dagger$ metric over all test samples.}\\
\multicolumn{9}{l}{$^*$ average over the metric  of the 5 test tiles.}\\
\end{tabular}
}
\end{table*}

\begin{table*}
\caption{Zurich Summer dataset: numerical results on the five test
  tiles. Each pair of columns illustrates the change of performance
  between the unaries (CNN) and the {spatial} model using them as a
  base {probabilistic input in the unary potential}. The CRF is based
  on~\cite{Boykov2001}.
  \label{tab:num}}
\setlength{\tabcolsep}{\mysize}
\scriptsize{
\vspace{.1cm}
\begin{tabular}{c|ccl |ccl ||ccl |ccl ||ccl |ccl}
\cline{2-19}
      & \multicolumn{6}{c||}{Ov. Accuracy (OA)} & \multicolumn{6}{c||}{Kappa ($\kappa$)} &  \multicolumn{6}{c}{Av. Accuracy (AA)}\\\hline
Tile  & \multicolumn{3}{c|}{Pixels} & \multicolumn{3}{c||}{Regions} & \multicolumn{3}{c|}{Pixels} & \multicolumn{3}{c||}{Regions} & \multicolumn{3}{c|}{Pixels} & \multicolumn{3}{c}{Regions} \\
\#    & {CNN} & {CRF}& {2L$\lightning$CRF} & {CNN} & {CRF}& {2L$\lightning$CRF} & {CNN} & {CRF}& {2L$\lightning$CRF} & {CNN} & {CRF}& {2L$\lightning$CRF} &  {CNN} & {CRF}& {2L$\lightning$CRF} &  {CNN}& {CRF} & {2L$\lightning$CRF} \\
\hline
16 & 90.7	&90.7&91.0 	&87.0 &87.0&92.3  	&0.87 &0.87&0.87&		0.82 &0.82& 0.89	&73.2&73.2&73.2	&69.1 &69.0& 78.1 \\
17 & 90.2	&90.2&90.4 &88.5	&88.5&93.0 	&0.87&0.87&0.87&	0.85 &0.85& 0.91&69.5&69.4&69.5&70.0&70.0& 71.9 \\
18 & 91.1	&91.1&91.2 &90.7	&90.7&92.7 	 &0.87&0.87&0.87&	0.86 &0.86&0.89&86.9&86.9&87.1& 85.3&85.3&88.0 \\
19 & 90.4	&90.4&90.4&87.7	&87.7&91.4	 &0.87&0.87&0.87&	 0.84 &0.84&0.89&91.1&91.1&91.2&87.2&87.2&  92.9 \\
20 & 89.6	&89.7&89.9&88.0	&88.0&91.1&0.86&0.86&0.87	&0.84&0.84& 0.89 &76.0&76.0&76.2&74.1&74.1&77.3 \\
\hline
Avg.$^\dagger$ & 90.3 &90.3&90.5 \green{(+0.2)}&	88.1&88.1&92.0	\green{(+3.9)} &0.88&0.88&0.88& 0.85& 0.85& 0.90 \green{(+0.05)} &77.7&77.7&77.8 \green{(+0.1)}&76.5&76.5& 79.3 \green{(+2.8)} \\
Ov.$^*$ & 90.4	&90.4&90.6 \green{(+0.2)}&88.3	&88.4&92.1 \green{(+3.7)} &0.87&0.87&0.87& 0.84&0.84&0.89 \green{(+0.05)} &79.3&79.4&79.4 &77.1&77.1&81.6 \green{(+4.5)} \\
\hline
\multicolumn{9}{l}{$^\dagger$ metric over all test samples.}\\
\multicolumn{9}{l}{$^*$ average over the metric  of the 5 test tiles.}\\
\end{tabular}
}
\end{table*}

\noindent \textbf{Random forests vs. CNNs.} Table~\ref{tab:num}
presents the numerical results of the same dataset, this time obtained
by a CNN predicting single labels per patch. The first striking
observation is that all the figures of merit have a sharp improvement
of about 10-15\% in overall accuracy (OA), 10-18 $\kappa$ points or
8-10\% in {AA}. Certainly the random forest results could have been
improved by a more in depth research on feature engineering of the
input space. {However, considering that the CNN learns
  end-to-end, we only devoted efforts in finding the appropriate
  architecture training well.} This is in line with observations in
several other recent papers. 

\begin{figure*}[!t]
\begin{tabular}{lcc|c|cc}
&&& Pixel level & \multicolumn{2}{c}{Region level} \\
\rotatebox{90}{\#16} &
\includegraphics[width=.17\textwidth]{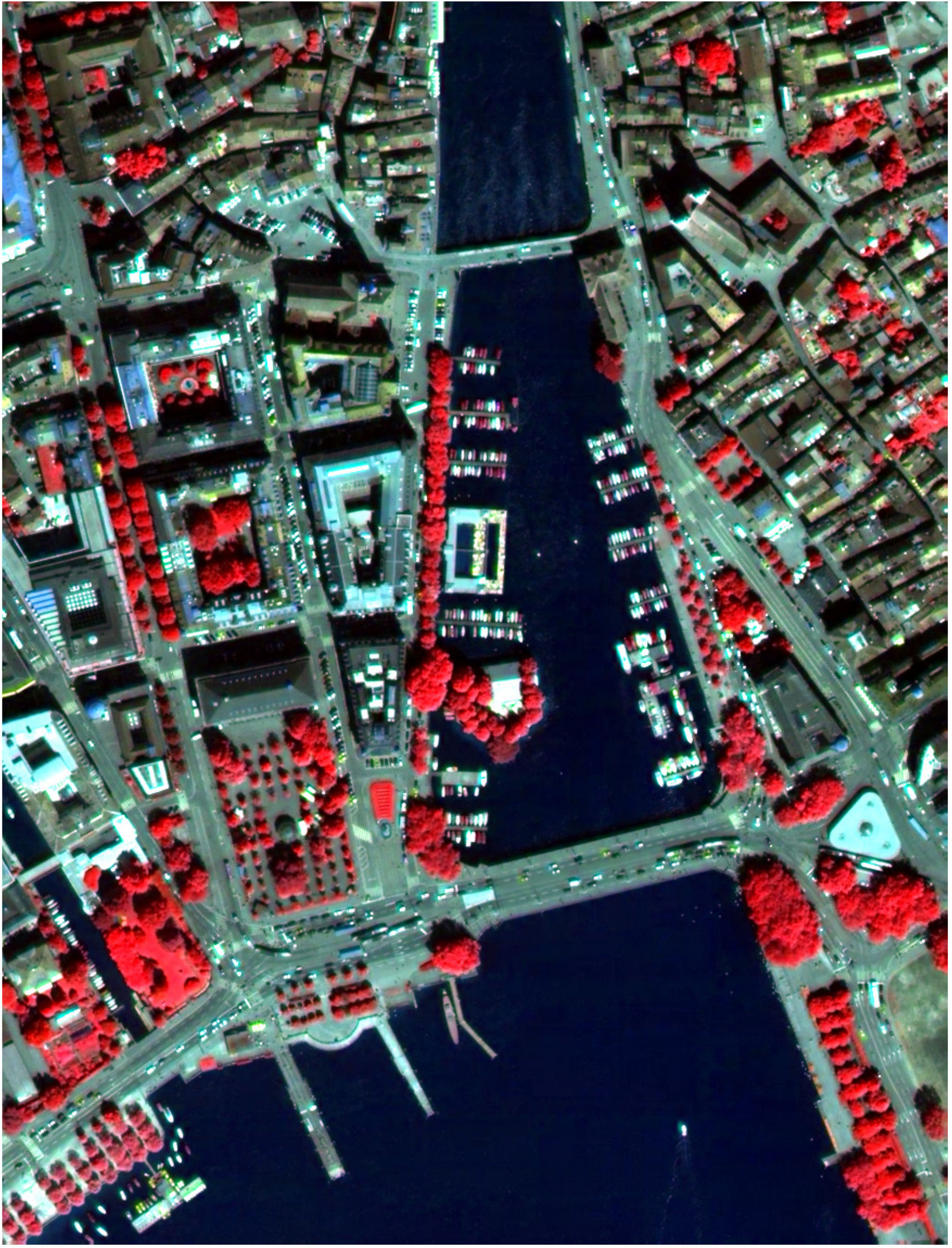}&
\includegraphics[width=.17\textwidth]{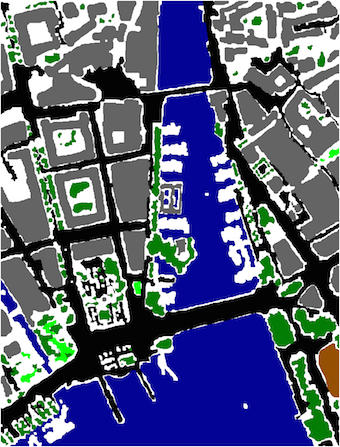}&
\includegraphics[width=.17\textwidth]{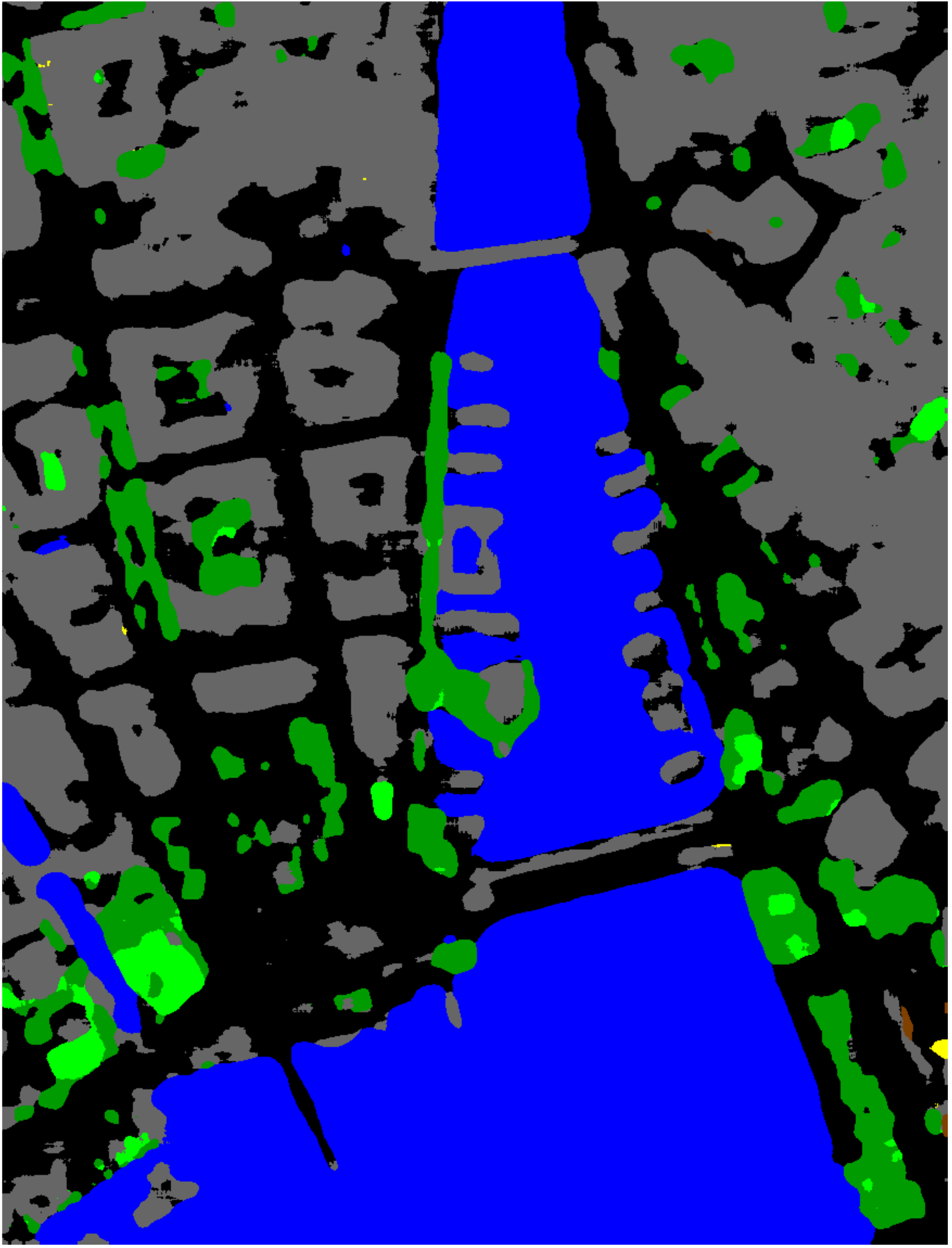}&
\includegraphics[width=.17\textwidth]{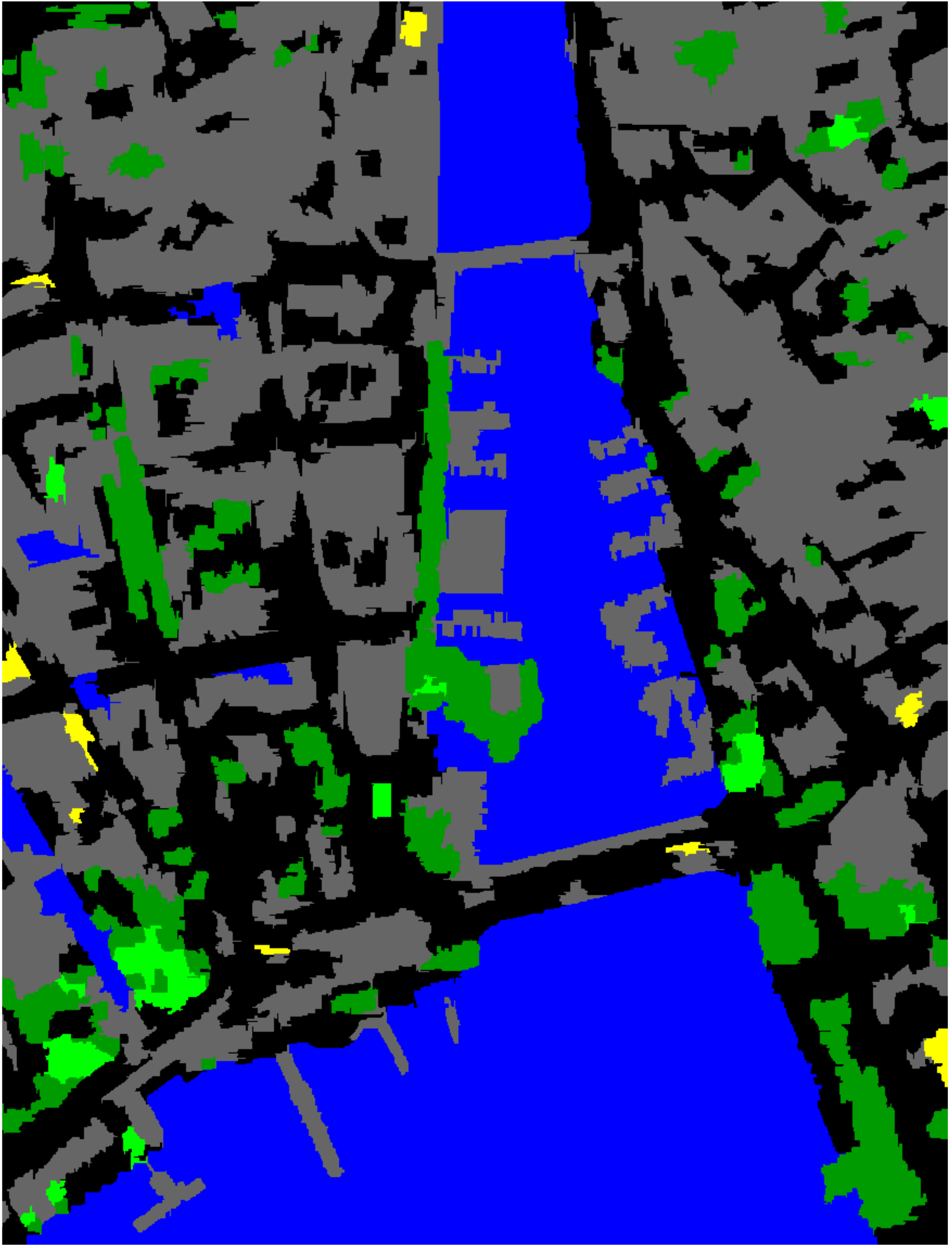}&
\includegraphics[width=.17\textwidth]{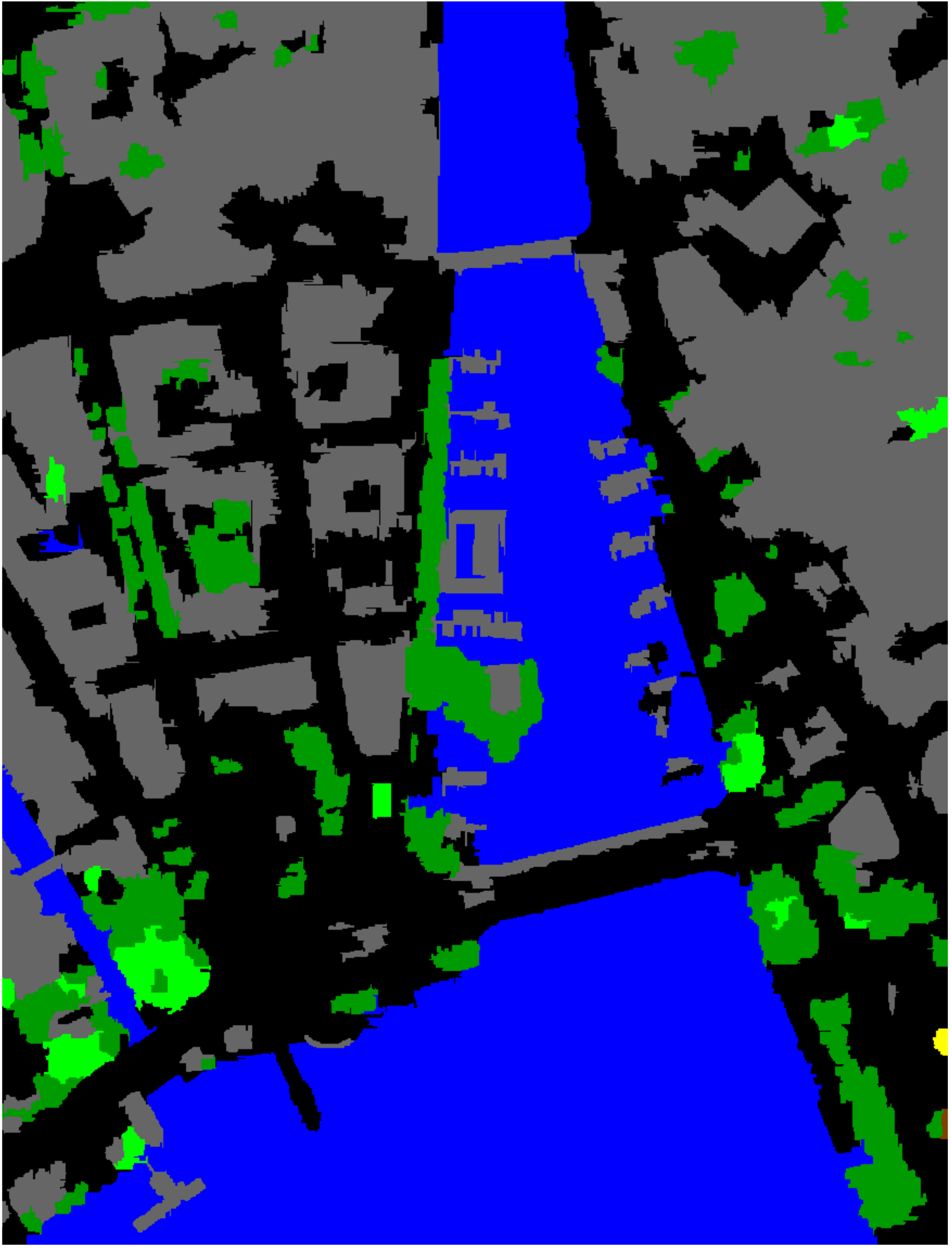}\\[-0.5mm]
\rotatebox{90}{\#17} &
\includegraphics[width=.17\textwidth]{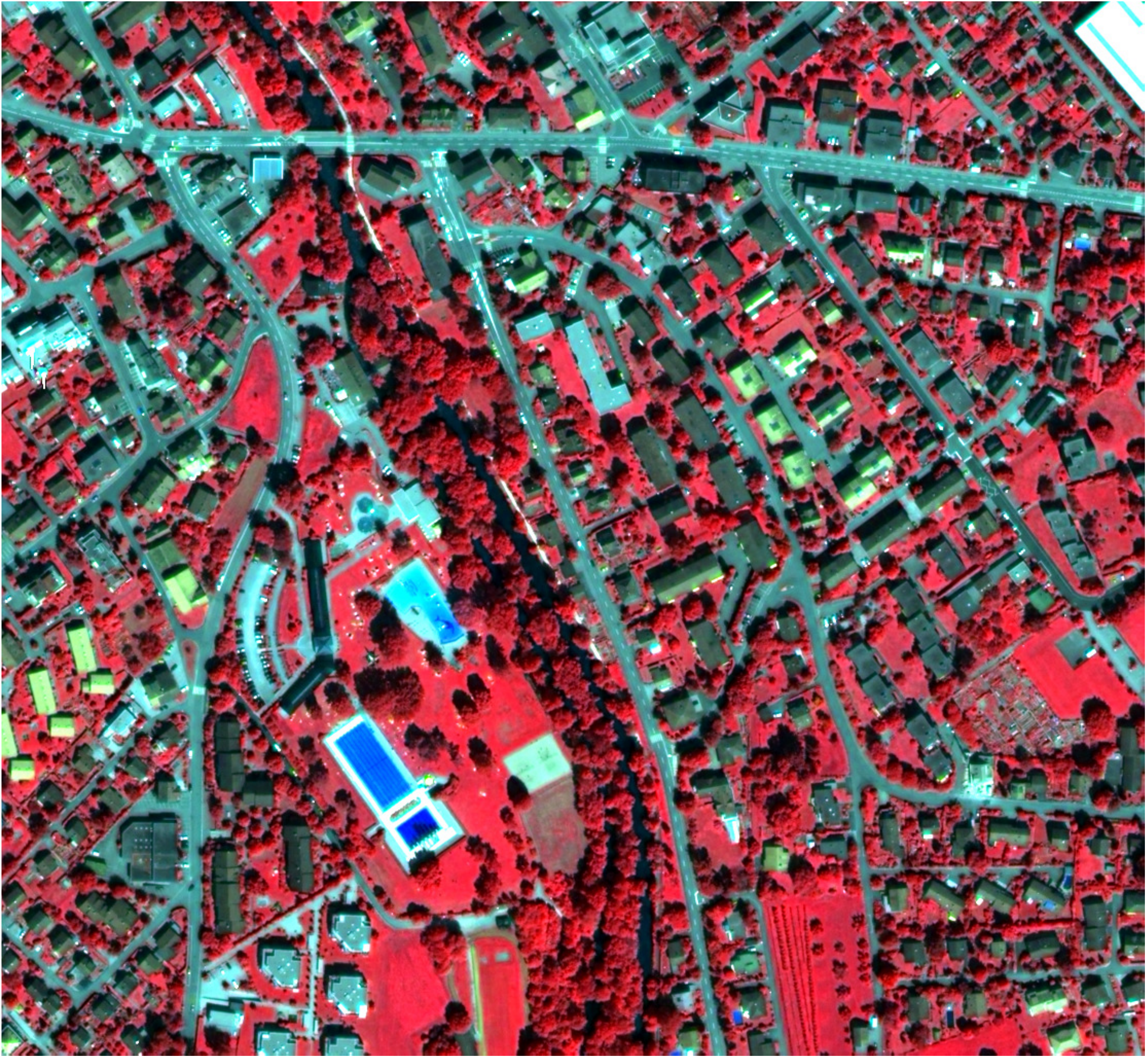}&
\includegraphics[width=.17\textwidth]{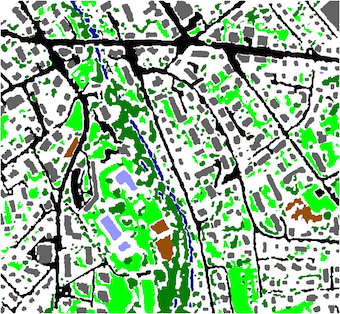}&
\includegraphics[width=.17\textwidth]{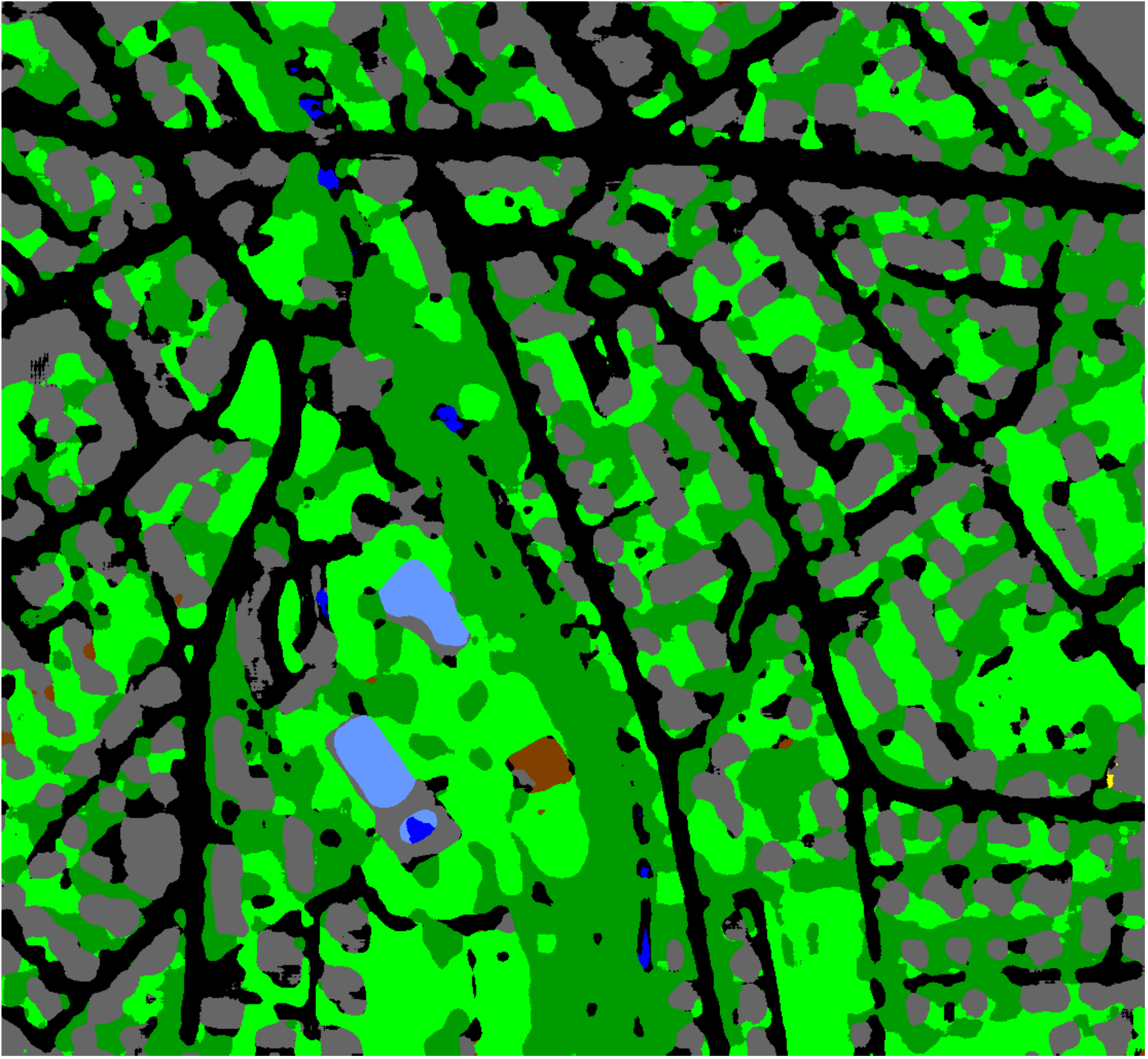}&
\includegraphics[width=.17\textwidth]{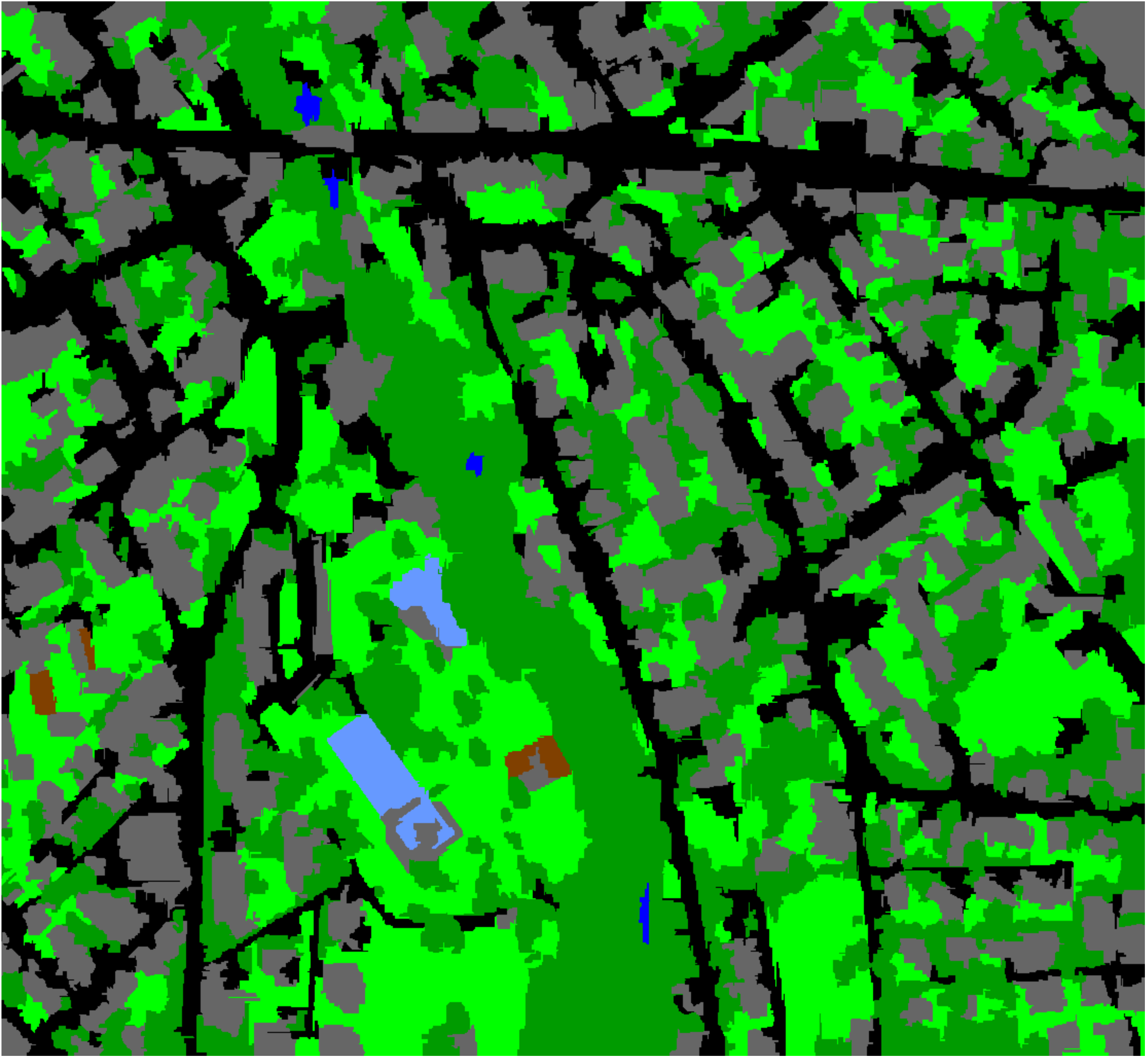}&
\includegraphics[width=.17\textwidth]{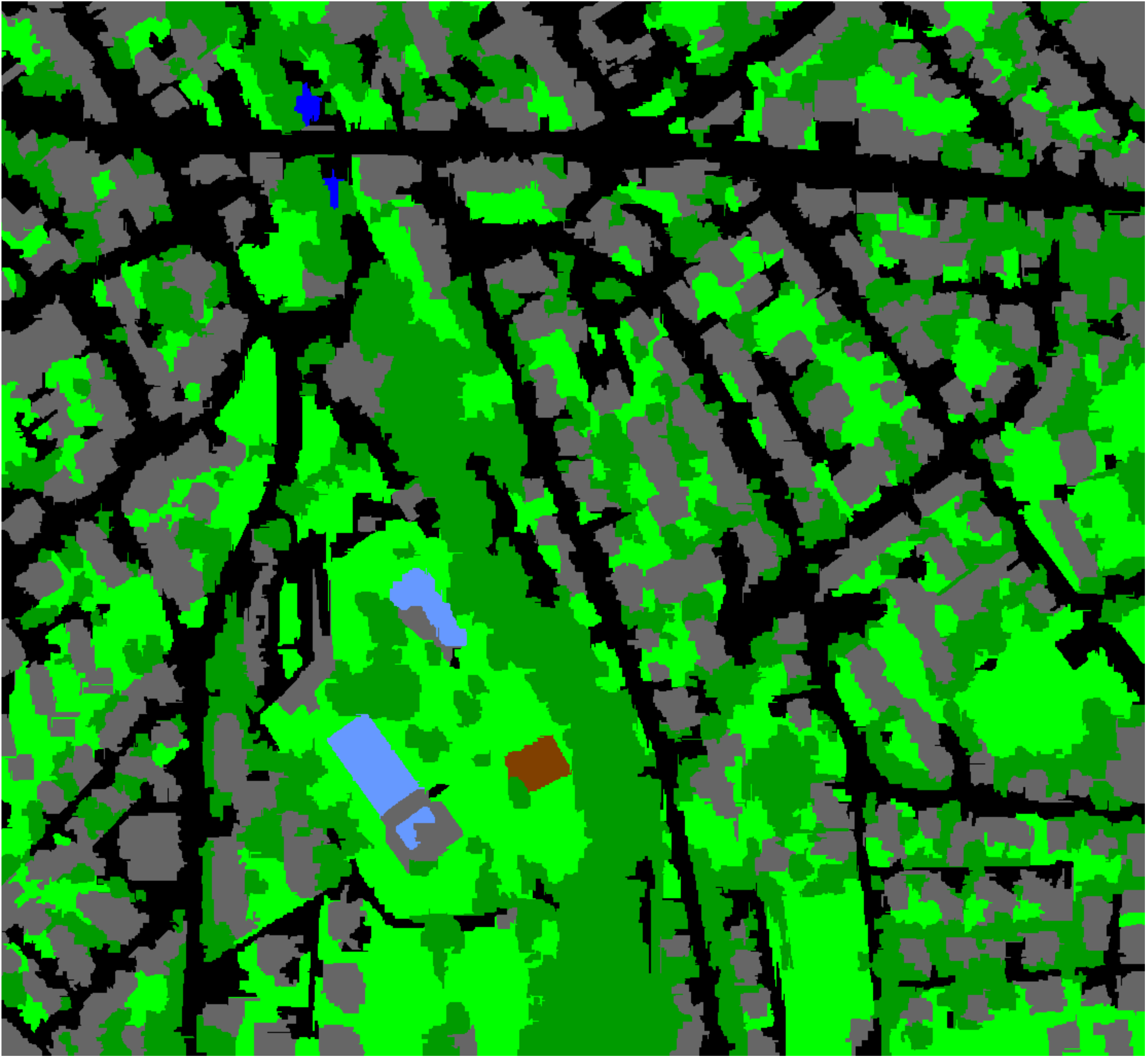}\\[-0.5mm]
\rotatebox{90}{\#18} &
\includegraphics[width=.17\textwidth]{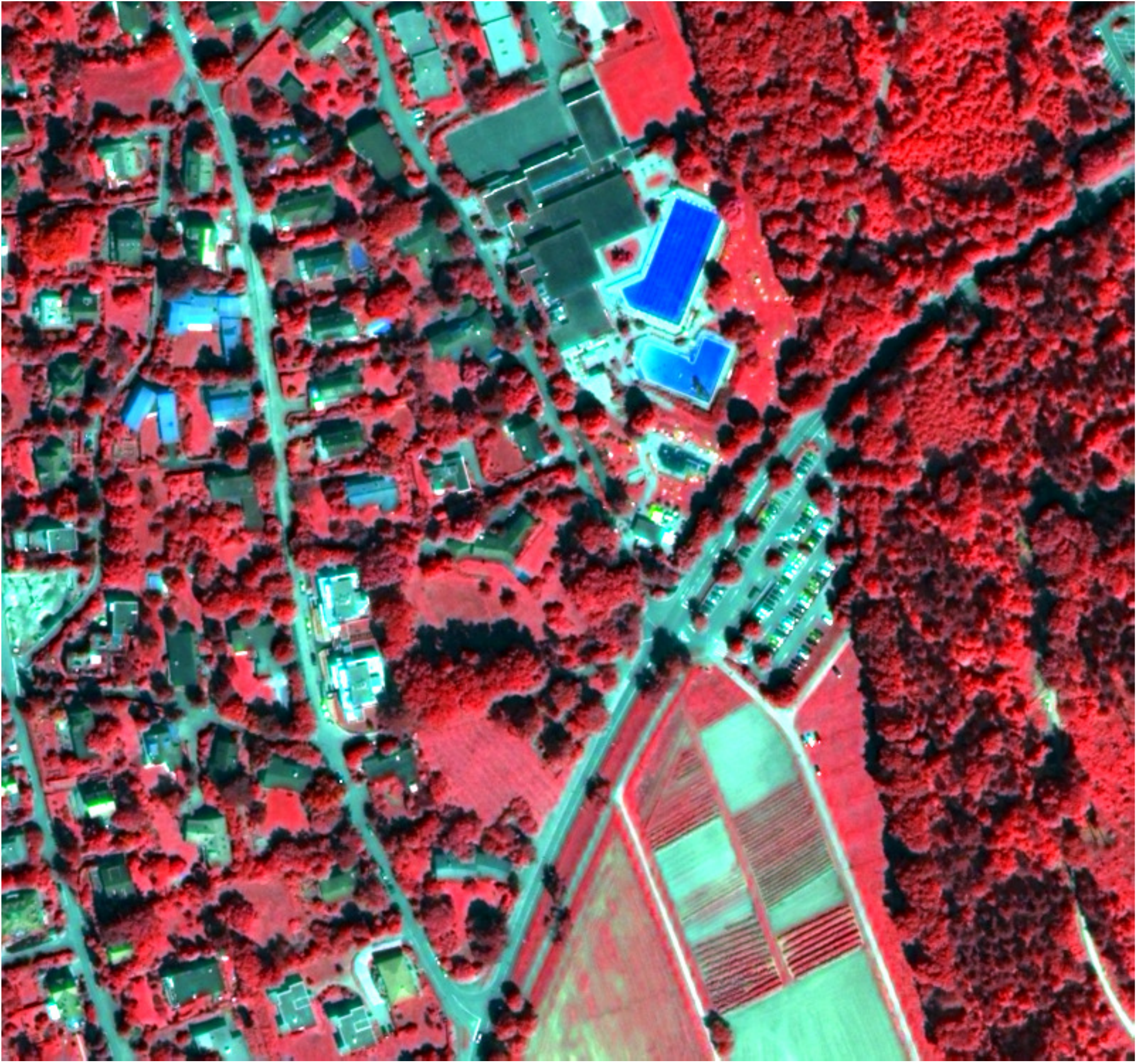}&
\includegraphics[width=.17\textwidth]{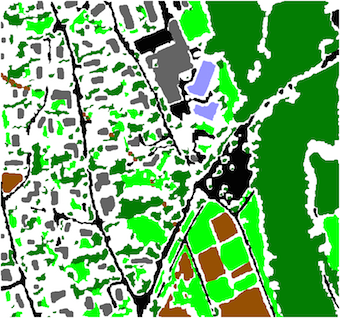}&
\includegraphics[width=.17\textwidth]{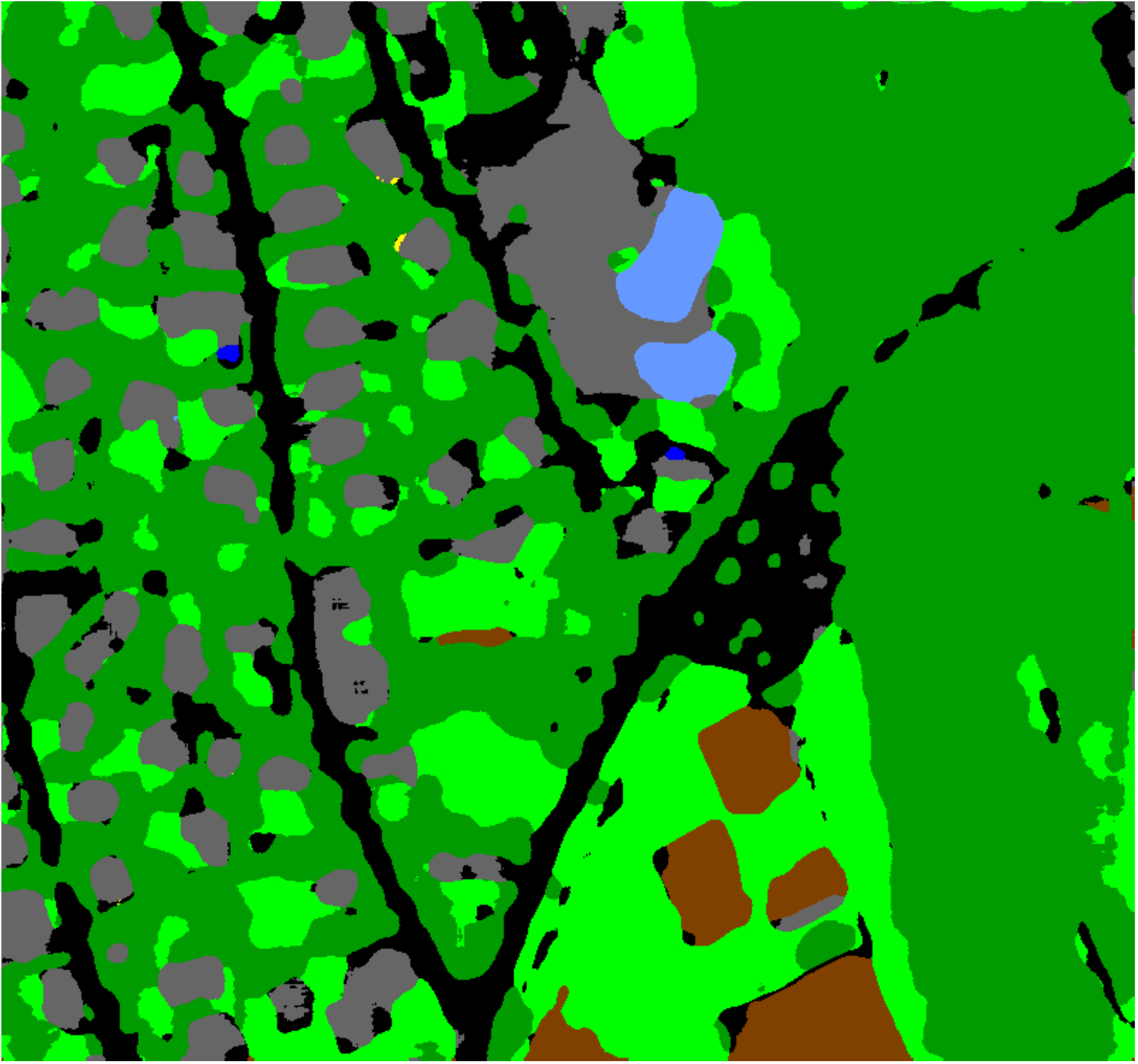}&
\includegraphics[width=.17\textwidth]{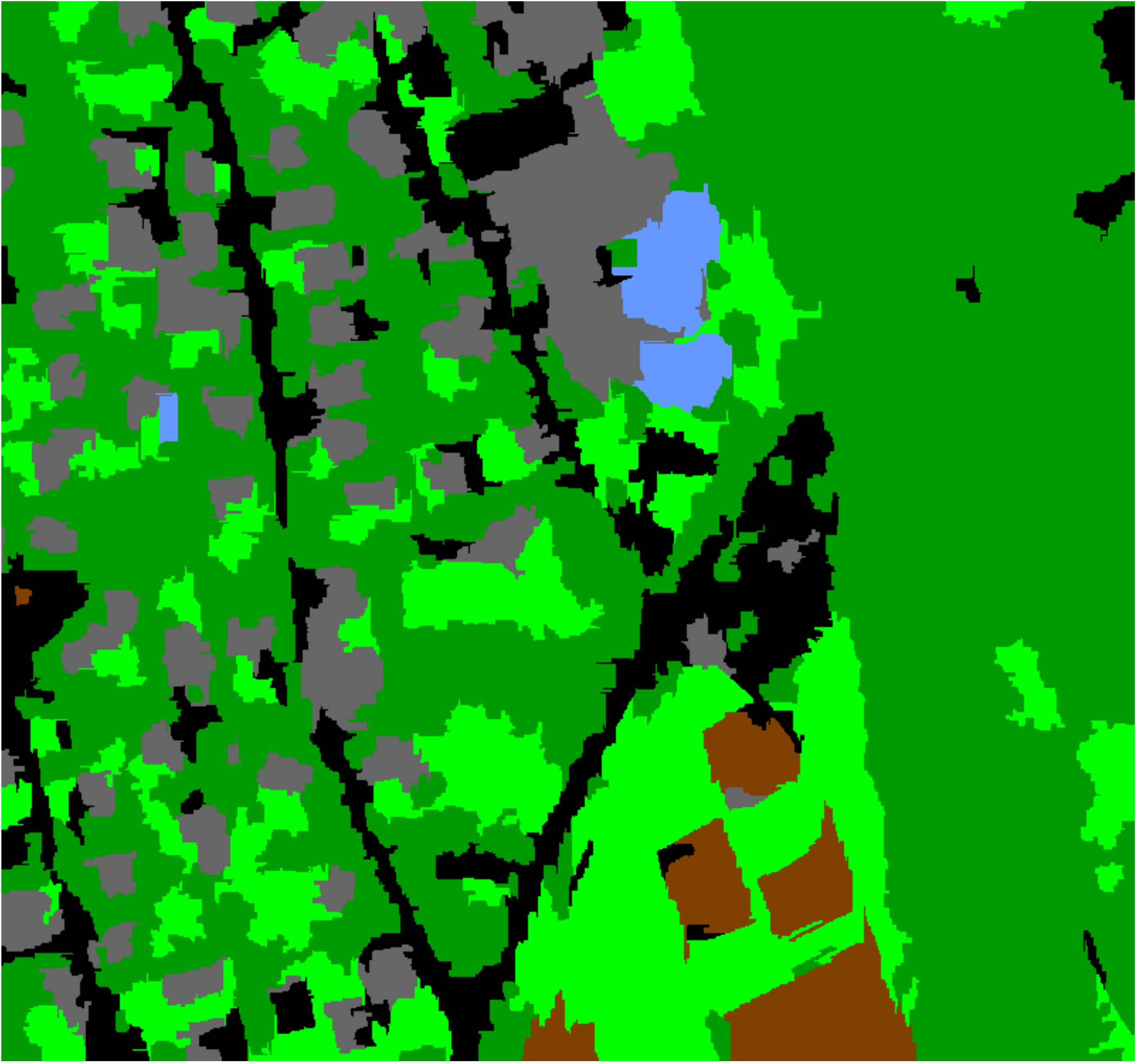}&
\includegraphics[width=.17\textwidth]{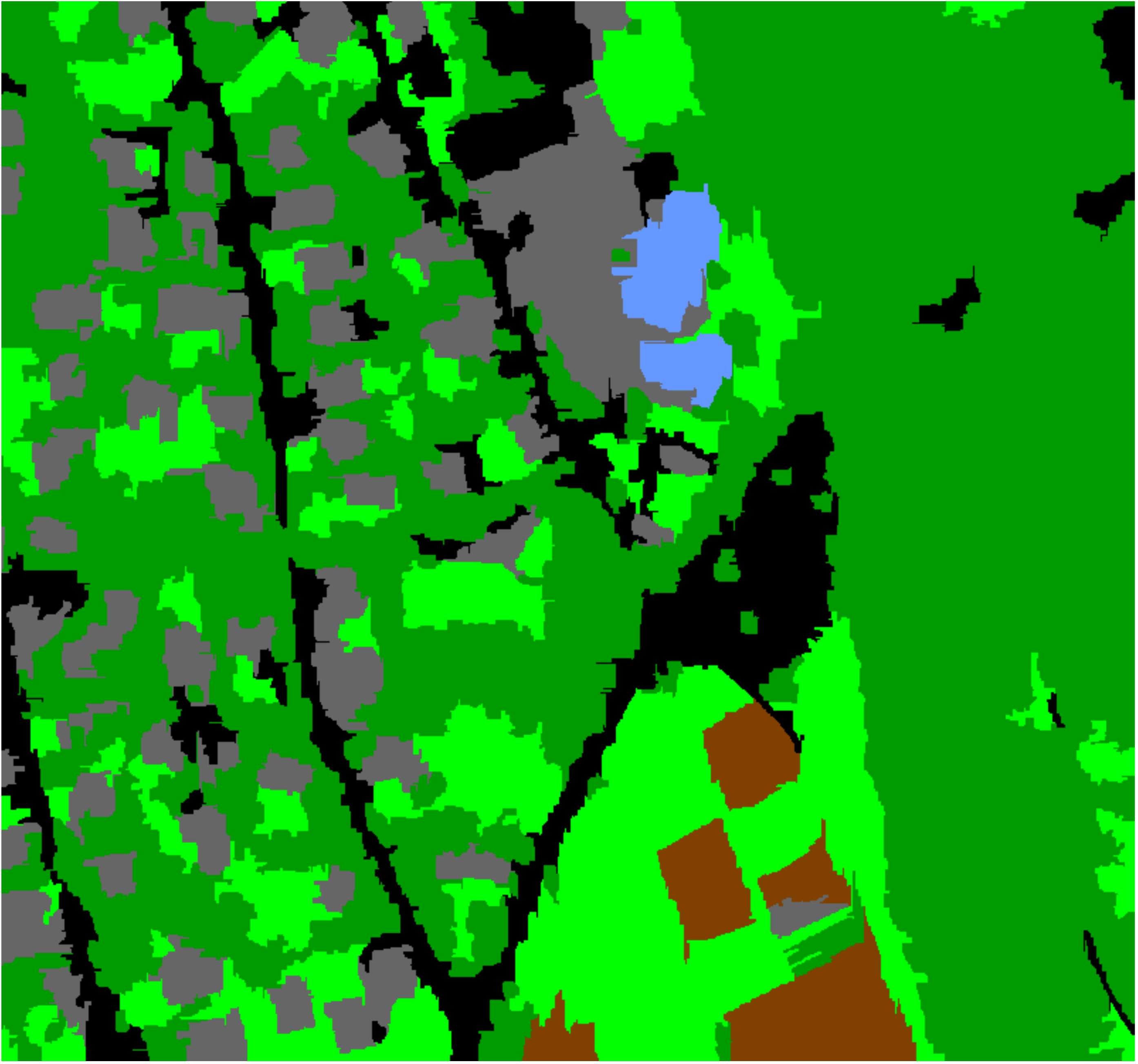}\\[-0.5mm]
\rotatebox{90}{\#19} &
\includegraphics[width=.17\textwidth]{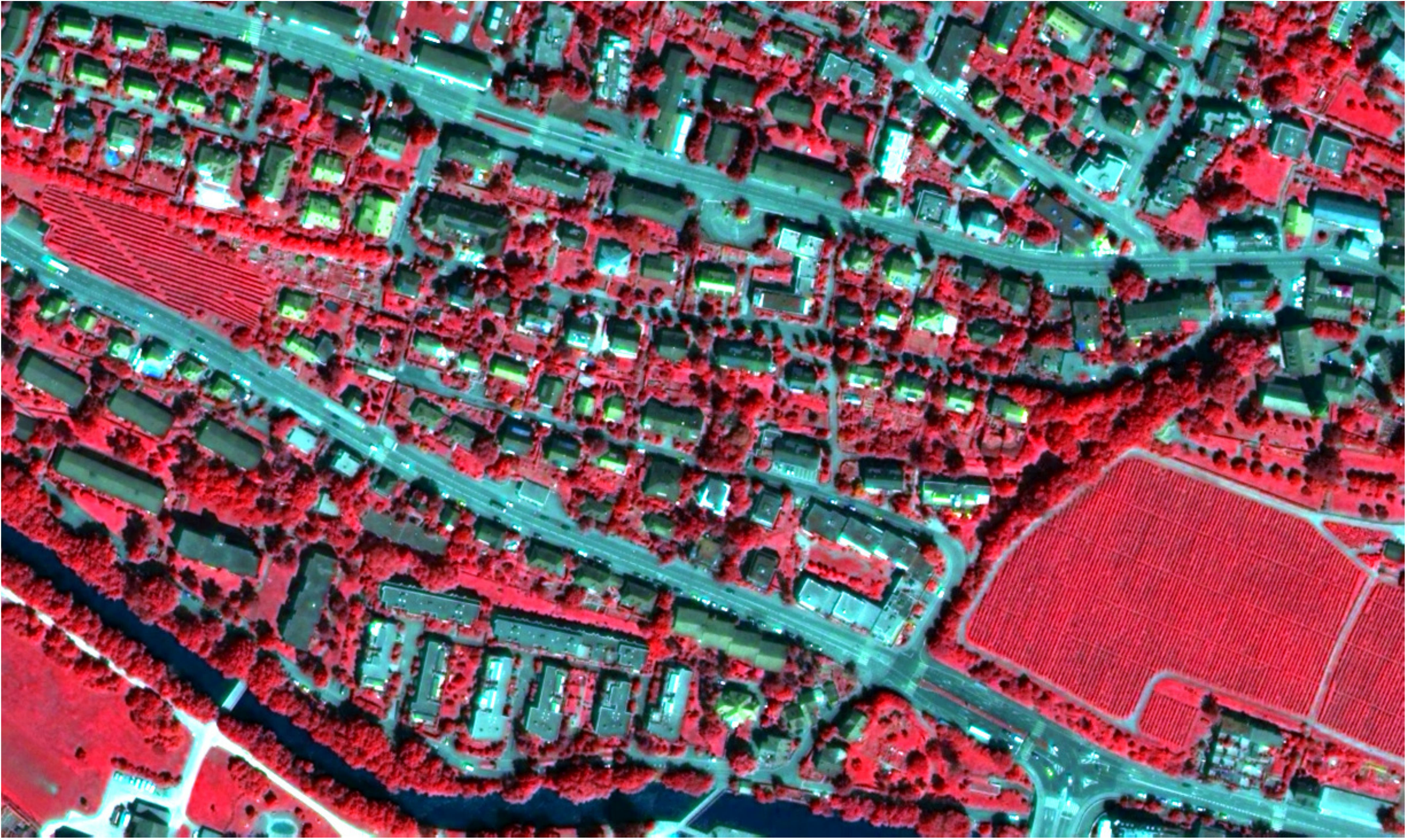}&
\includegraphics[width=.17\textwidth]{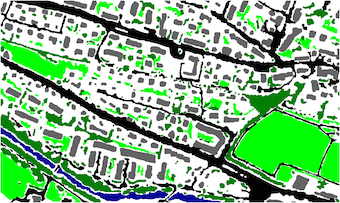}&
\includegraphics[width=.17\textwidth]{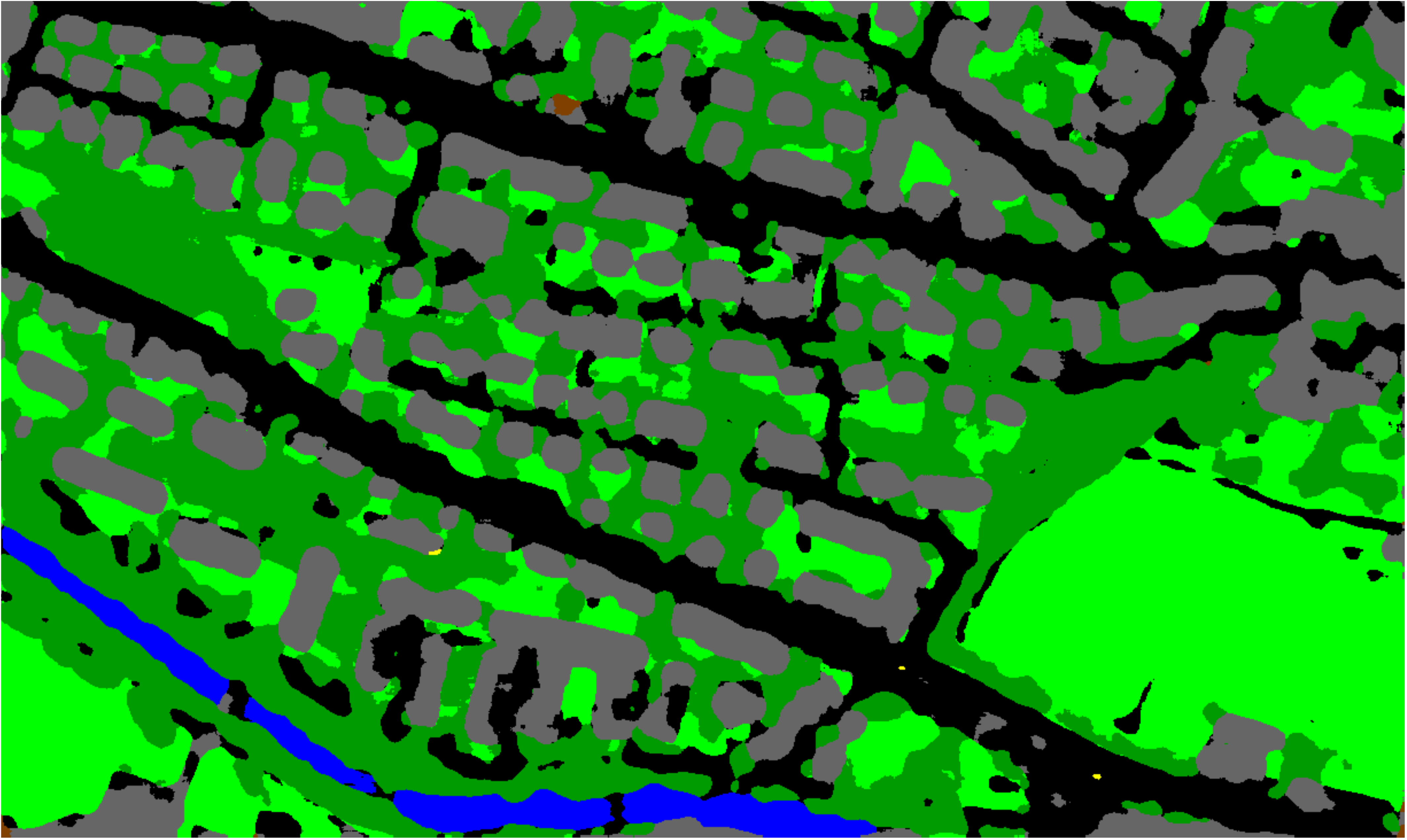}&
\includegraphics[width=.17\textwidth]{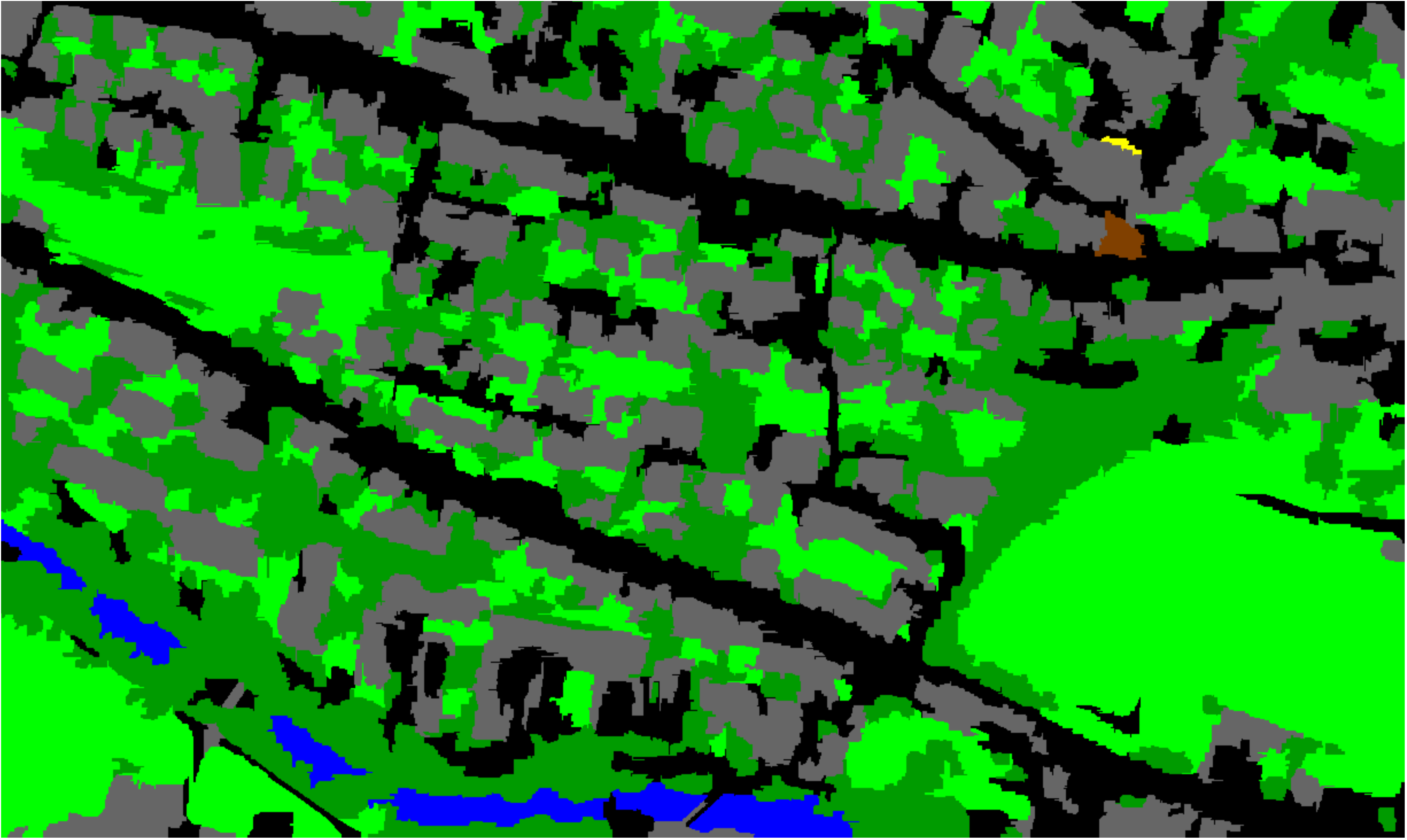}&
\includegraphics[width=.17\textwidth]{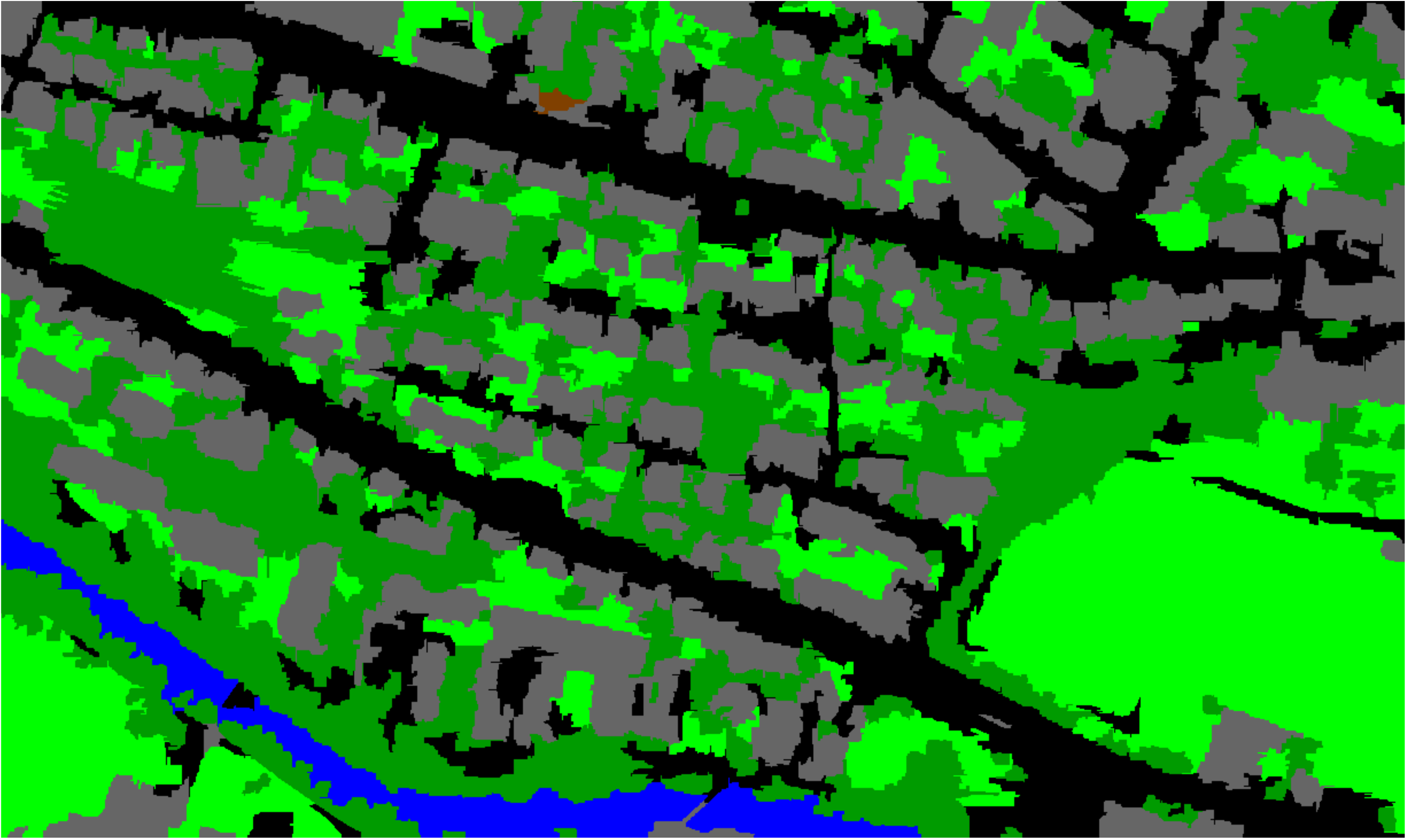}\\[-0.5mm]
\rotatebox{90}{\#20} &
\includegraphics[width=.17\textwidth]{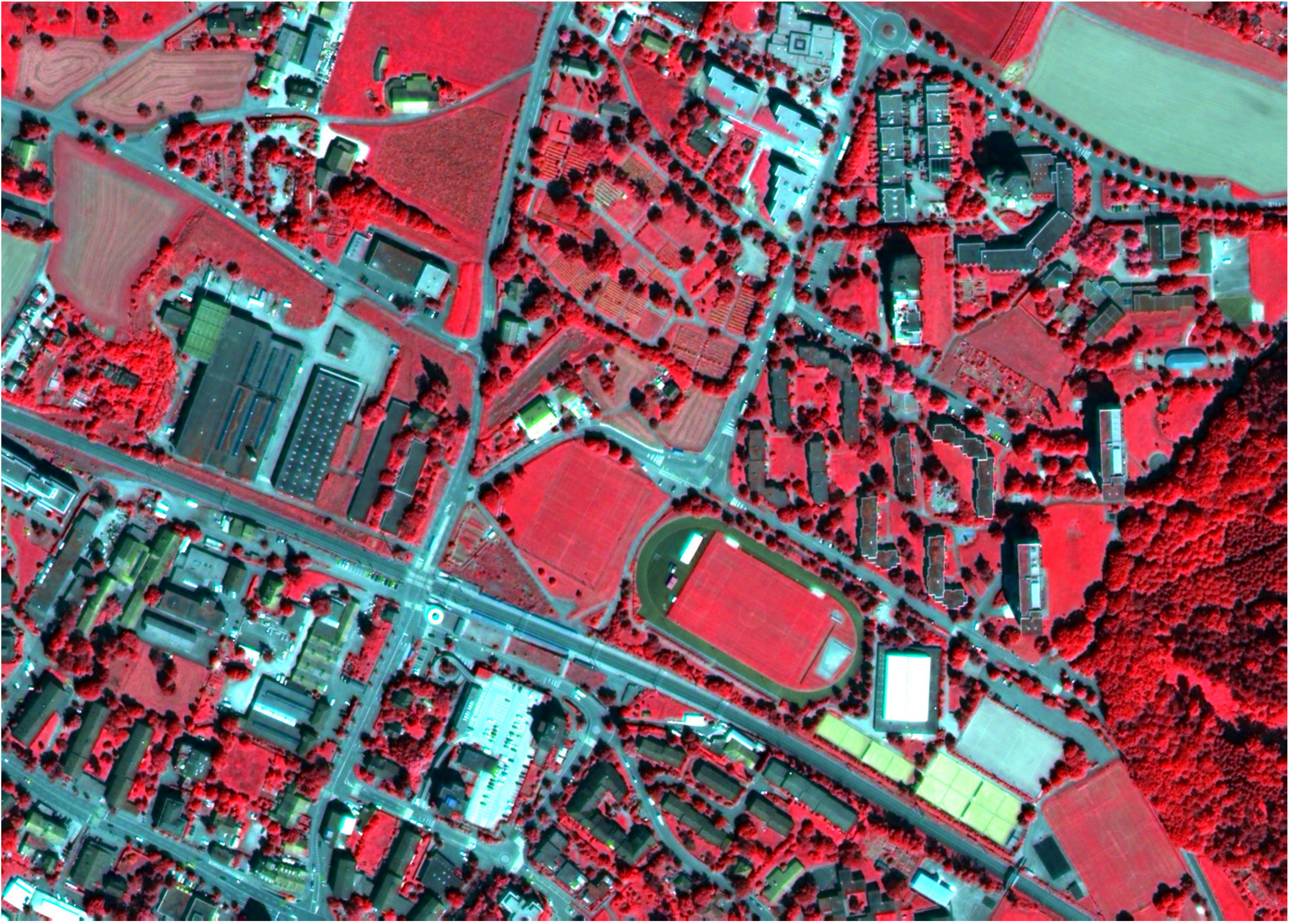}&
\includegraphics[width=.17\textwidth]{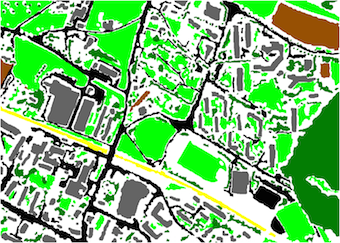}&
\includegraphics[width=.17\textwidth]{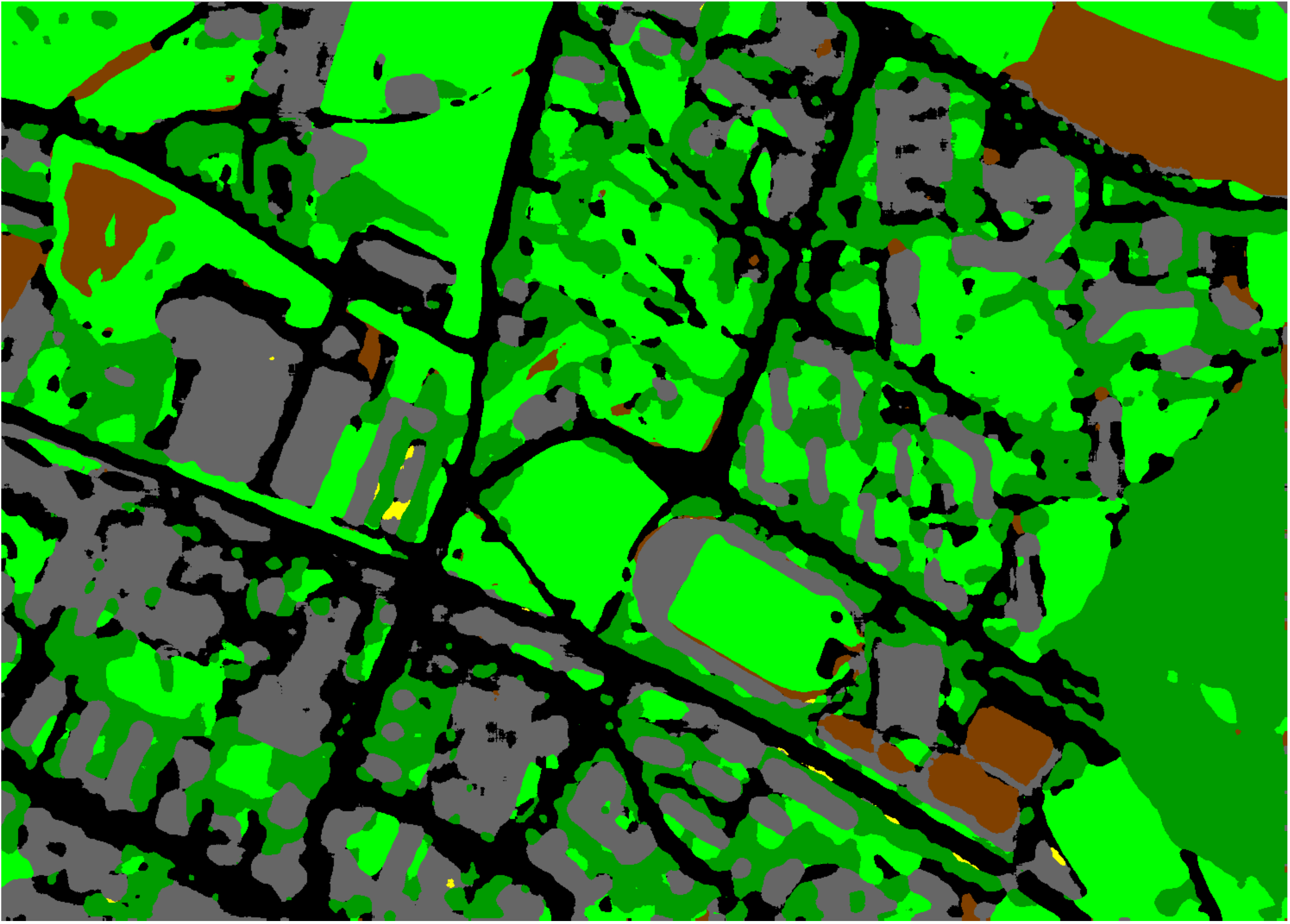}&
\includegraphics[width=.17\textwidth]{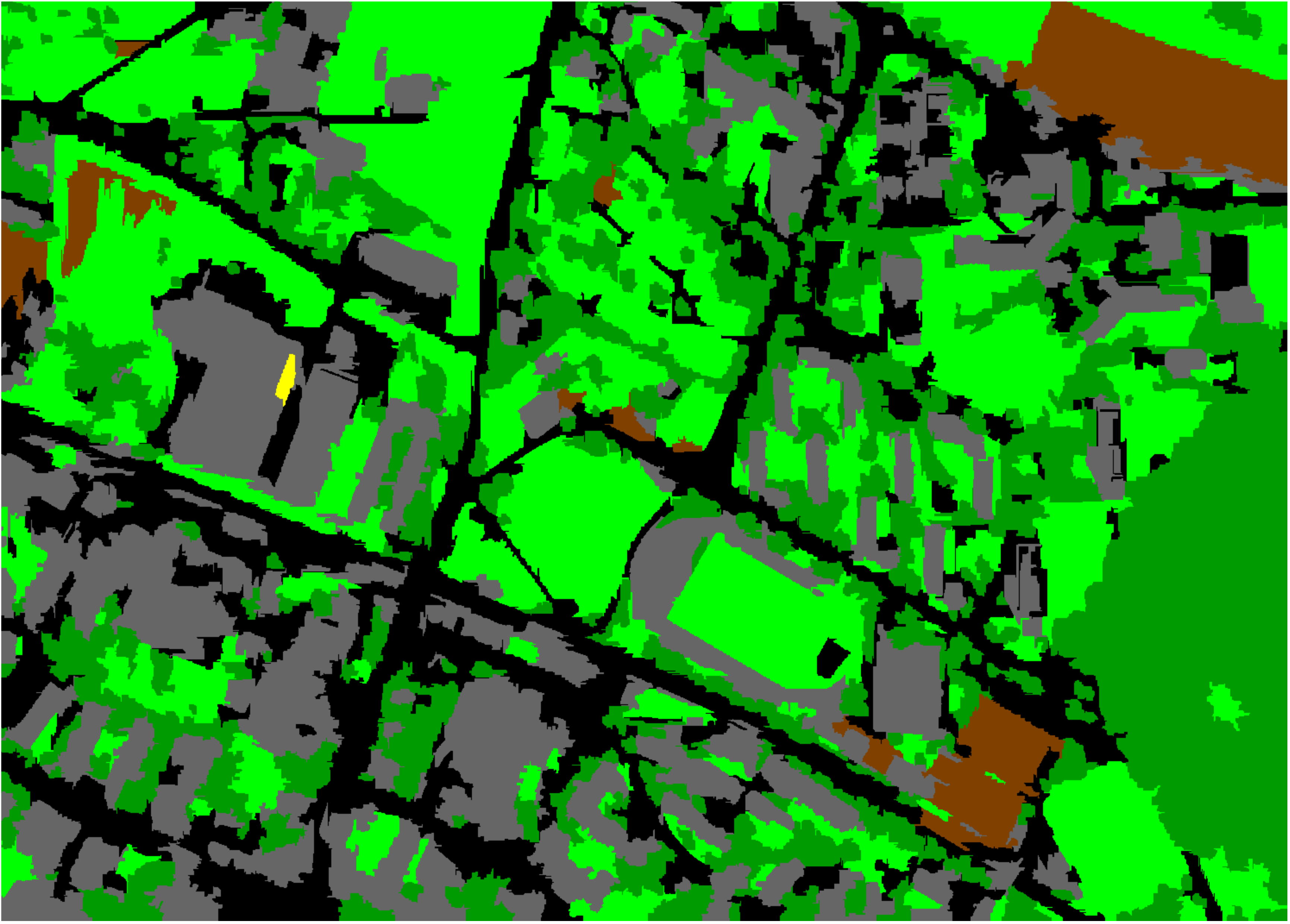}&
\includegraphics[width=.17\textwidth]{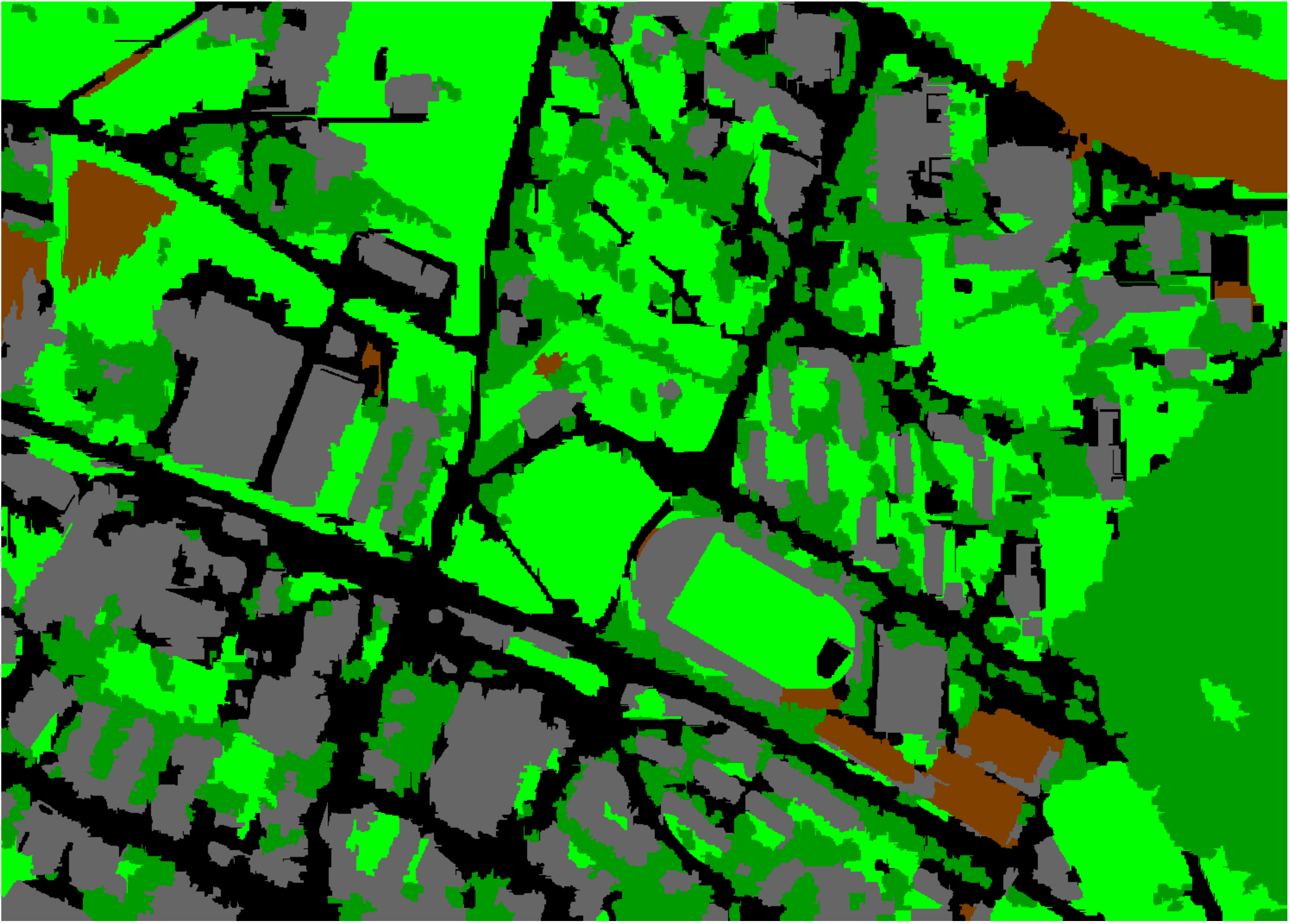}\\[-0.5mm]
& 
(a) image & (b) ground truth & (c) CNN & 
(d) CNN & (e) 2L$\lightning$CRF \\[-0.5mm]
\end{tabular}
\caption{Zurich Summer dataset: results on the five test images. (a) original image; (b) ground truth; (c) CNN, pixel-based; 
(d) CNN, region-level; 
(e) 2L$\lightning$CRF, region-level. For the accuracies of the single maps, please refer to Tab.~\ref{tab:num} (color legend: \textcolor{resi}{residential}, \textcolor{street}{street}, \textcolor{tr}{trees}, \textcolor{mea}{meadows}, \textcolor{rail}{railway}, \textcolor{wat}{water}, \textcolor{pool}{swimming pools}, \textcolor{bare}{bare soil}). \label{fig:resMaps}}
\vspace*{-3mm}
\end{figure*}

\noindent \textbf{CNNs.} Considering the role of the proposed
2L$\lightning$CRF {when using} unary potentials from a CNN, {we observe a general
improvement of the accuracy figures}, although it is less striking
 than in the previous case of random forests, in particular on pixels
 results. The reason lies in the high accuracy of the initial result,
 for which the unary scores are sharp and spatially consistent (see
 the third column of Fig.~\ref{fig:resMaps}). This can be seen in the
 difference between the CNN results and those of the CRF {based
  on}~\cite{Boykov2001}, which considers only spatial interactions for
each lattice separately. The 2L$\lightning$CRF model only needs to
correct for inconsistent {labelings} among the two layers, which
results in small increases in the accuracies. On the contrary, the
high accuracy and consistency of the unary scores allows to correct
for several misclassifications at the region level, for which the
pooling of the unary scores resulted in erroneous region scores, in
particular in the case of regions with high entropy of the posterior
probability at the pixel level. For instance, the road network is
poorly reconstructed in tile \#16 in Fig.~\ref{fig:resMaps}, as
regions composing it are often incorrectly classified as buildings. On
the contrary, they are assigned to the roads class by
2L$\lightning$CRF, since the evidence from the pixel representation is
strong enough to {influence} the response at the region level, thus
leading to a map that is correct (by the pixel response) and sharp (by
the region topology). Another example is in tile \#19, where the river
is correctly recovered by 2L$\lightning$CRF, while the original region
results were providing a mix between water, trees, and roads.

\begin{figure}[!t]

\begin{tabular}{lcc}
& \multicolumn{2}{c}{Pixels} \\

\rotatebox{90}{\hspace{1cm} OA}&
\includegraphics[width = 3.5cm]{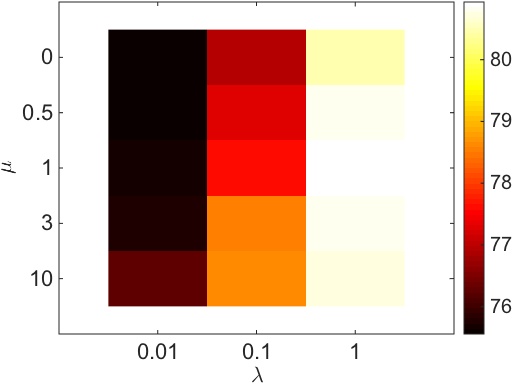} & 
\includegraphics[width = 3.5cm]{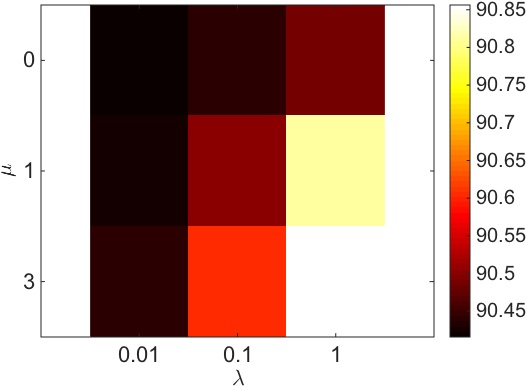} \\

\rotatebox{90}{\hspace{1cm}  AA}&
\includegraphics[width = 3.5cm]{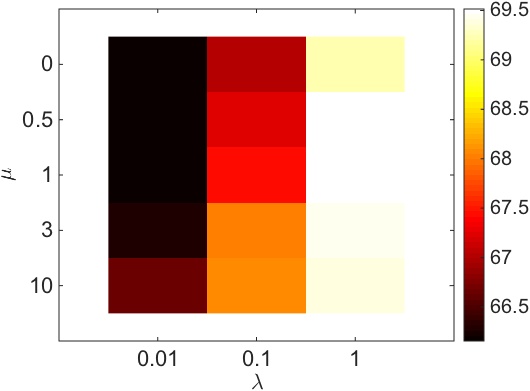} & 
\includegraphics[width = 3.5cm]{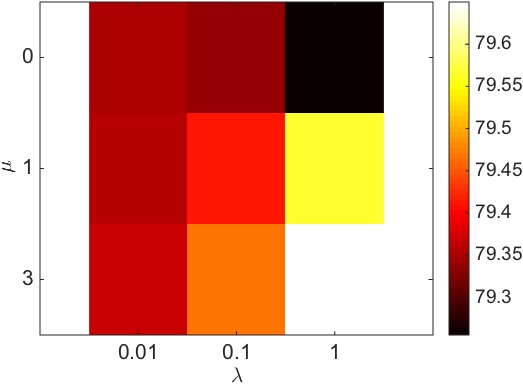} \\
\hline

& \multicolumn{2}{c}{Regions} \\
\rotatebox{90}{\hspace{1cm}  OA}&
\includegraphics[width = 3.5cm]{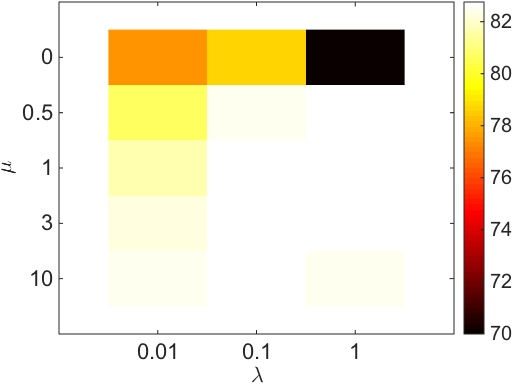} & 
\includegraphics[width = 3.5cm]{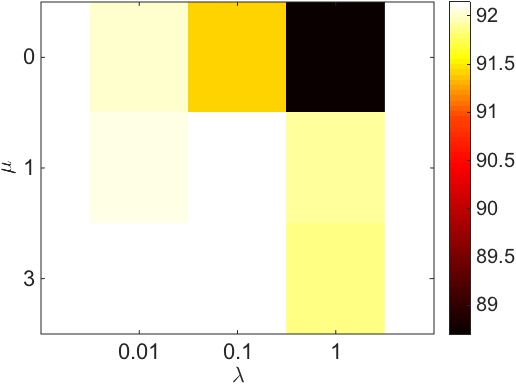} \\
\rotatebox{90}{\hspace{1cm}  AA}&
\includegraphics[width = 3.5cm]{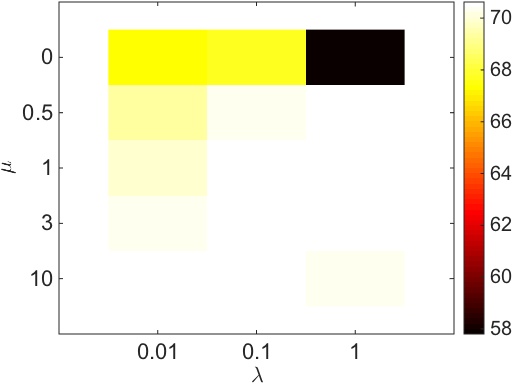} & 
\includegraphics[width = 3.5cm]{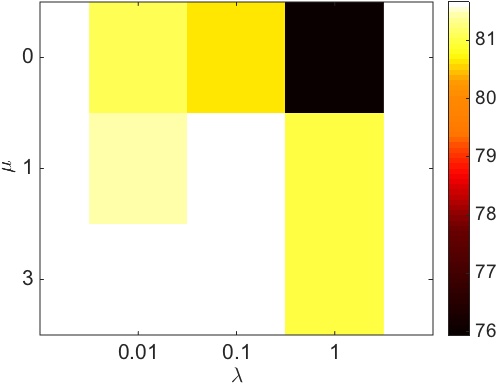} \\
& RF & CNN   \\
\end{tabular}
\caption{{Sensibility analysis} for the $\mu$ and $\lambda$ parameters in Eq.~\eqref{eq:E} and the Zurich Summer dataset.\label{fig:xval}}
\end{figure}

\noindent \textbf{Parameters analysis. } Figure~\ref{fig:xval}
  reports the {sensitivity analysis} for the $\mu$ and $\lambda$
  parameters, involved in Eq.~\eqref{eq:E}. As a reminder, $\mu$
  control the strength of the relations between the spatial supports
  and $\lambda$ the strength of the local spatial smoothing
  {within} each spatial support. In all experiments, the
  model not making any use of inter-support connections ($\mu=0$,
  which would roughly correspond to {optimizing} two
  {separate} CRFs) does not provide the best results,
  thus showing the interest of a joint model. This is particularly
  visible at the region level, where the evidence from the pixel level
  is able to {successfully revert many errors}, in
  particular for the RF classifier, {which often provides
    ambiguous class-likelihoods}. Also, using contrast sensitivity
  seems particularly important at the pixel level, while less at the
  region level: this is expected, since given the high resolution of
  the data, local inconsistencies at the pixel level are more frequent
  that noisy predictions at the region level.

\subsection{Zeebruges dataset}

\begin{figure*}[!t]
\includegraphics[width = .9\linewidth]{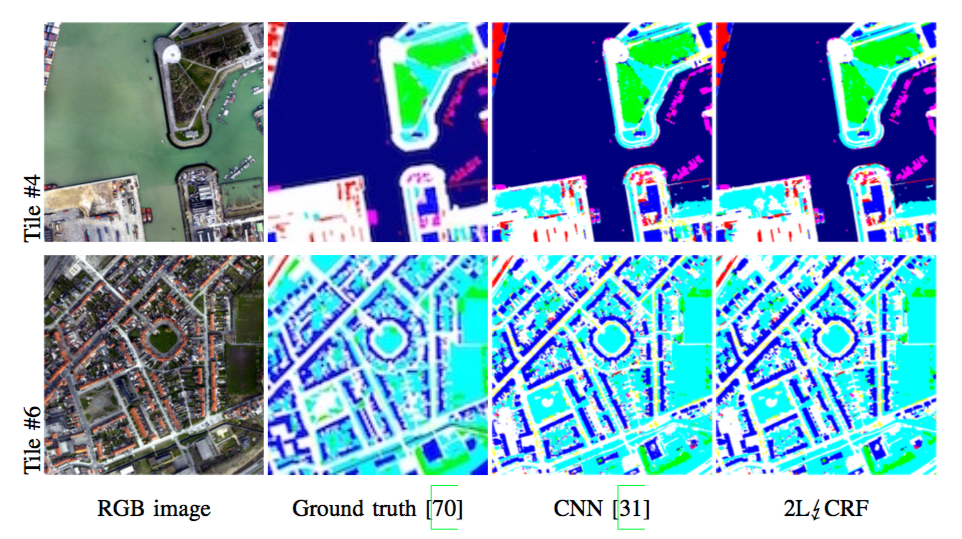}
\caption{Zeebruges dataset (\texttt{grss\_dfc\_2015}): results on the two test tiles. (a) original image; (b) ground truth; (c) CNN, pixel-based; 
(d) 2L$\lightning$CRF, pixel level. For the accuracies of the single maps, please refer to Tab.~\ref{tab:numZ} (color legend: \textcolor{Zimp}{impervious (white colored)}, \textcolor{Zwater}{water}, \textcolor{Zclutter}{clutter}, \textcolor{ZlowVeg}{low vegetation}, \textcolor{Zbuilding}{buildings}, \textcolor{Ztree}{trees}, \textcolor{Zboat}{boats}, \textcolor{Zcar}{cars}). The ground truth images are blurred as in~\cite{DFCA}, since they are undisclosed. \textbf{(Note: in this preprint we had to decrease the graphics resolution. For full resolution, please refer to the published version, or contact the authors)}.\label{fig:resMapsZ}}
\end{figure*}

\begin{figure}[!t]
\includegraphics[width = .9\linewidth]{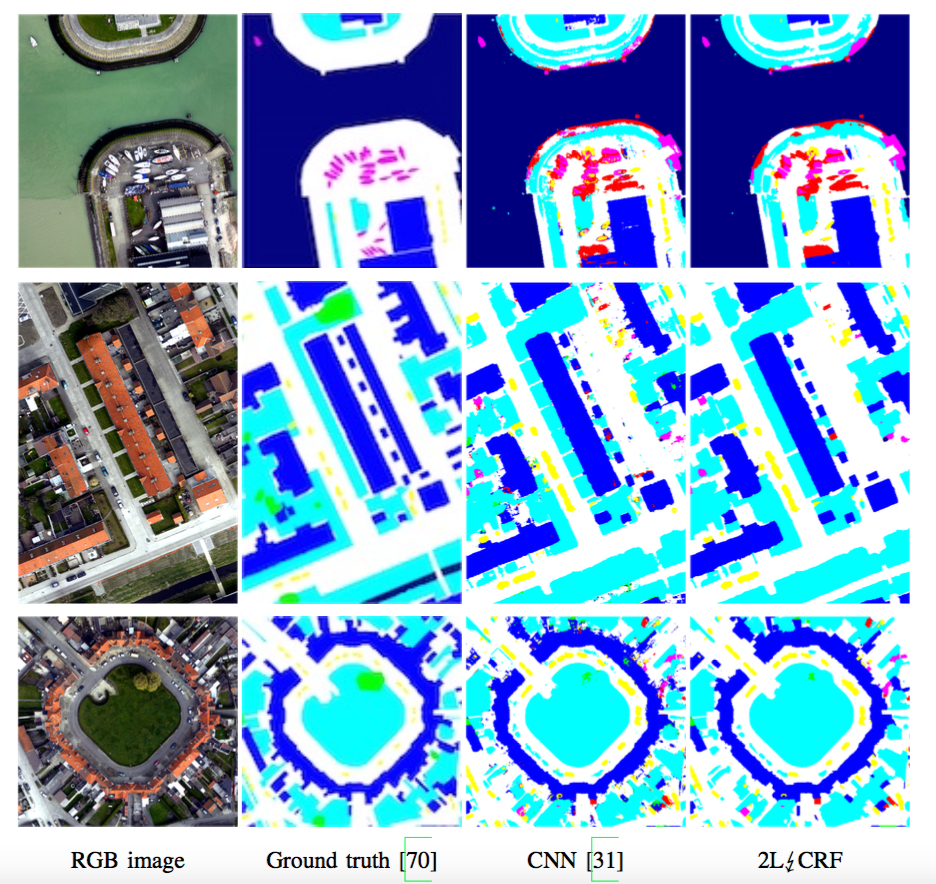}
\caption{Zeebruges dataset (\texttt{grss\_dfc\_2015}):  zoomed results on details of the two test tiles of Fig.~\ref{fig:resMapsZ}. (a) original image; (b) ground truth; (c) CNN, pixel-based; 
(d) 2L$\lightning$CRF, pixel level. For the accuracies of the single maps, please refer to Tab.~\ref{tab:numZ} (color legend: \textcolor{Zimp}{impervious (white coloured)}, \textcolor{Zwater}{water}, \textcolor{Zclutter}{clutter}, \textcolor{ZlowVeg}{low vegetation}, \textcolor{Zbuilding}{buildings}, \textcolor{Ztree}{trees}, \textcolor{Zboat}{boats}, \textcolor{Zcar}{cars}). The ground truth images are blurred as in~\cite{DFCA}, since they are undisclosed. \textbf{(Note: in this preprint we had to decrease the graphics resolution. For full resolution, please refer to the published version, or contact the authors)}. \label{fig:resMapsZZ}}
\end{figure}

In the case of the Zeebruges dataset, {we retrieve unary scores directly from the model of
  ~\cite{Mar17b}.} This model achieves results that compare favorably
to the {state-of-the-art methods} on this dataset.
It still suffers from some prediction noise leading to small
artifacts, as it can be seen in the maps in Fig.~\ref{fig:resMapsZ}
for the entire tiles and in Fig.~\ref{fig:resMapsZZ} for details. This
is certainly due to the increased spatial resolution (and therefore
complexity) of the problem, since we are working, compared to the
Zurich Summer dataset, on a ten times higher spatial resolution and
without an infrared channel.

Numerical results are reported in Tab.~\ref{tab:numZ}: as a first
observation, the standard CRF approach seems to improve the results
only marginally, which can be explained by two factors: first, and as
for the previous dataset, the CNN unaries are very sharp and confident
(even when misclassifying), making the change of a label very
unlikely. Secondly, the use of CRFs with up to pairwise nonzero clique
potentials rather than more sophisticated higher order CRFs (which was
necessary to keep the problem computationally feasible) made label
swaps improbable, since the pixel classifier was not much affected by
salt and pepper noise, but rather by larger spatial artifacts that could not
be corrected by looking only at the direct pixel neighborhood. This is
the reason why, for this dataset, 2L$\lightning$CRF greatly increases
performance at the pixel level: by letting the pixel lattice be
influenced by the region structure, where the scores are pooled per
region, the pixel CNN becomes aware of larger neighborhood structures
and can therefore correct for larger confident misclassifications at
pixel level (see the zooms in Fig.~\ref{fig:resMapsZZ} for some
examples). Such beneficial effect could have been maybe reinforced if
the region level CNN were a classifier trained specifically to predict
land cover at the region level (and not a set of region-pooled scores
over the pixel unaries), but this would have implied training a
separate CNN model for the region scale. {The consequent significant increase in memory and
computational resources would diminish -- in our opinion -- the
interest of the combined approach.}

\begin{table*}
\caption{NUmerical results on the Zeebruges dataset (\texttt{grss\_dfc\_2015}). Each pair of columns illustrates the
  change of performance between the unaries (CNN) and the spatial
  model using them as a base 
  in the unary potential. The CRF is based on~\cite{Boykov2001}.
\label{tab:numZ}}
\setlength{\tabcolsep}{\mysize}
\scriptsize{
\vspace{.1cm}
\begin{tabular}{c|ccl |ccl ||ccl |ccl ||ccl |ccl}
\cline{2-19}
      & \multicolumn{6}{c||}{Ov. Accuracy (OA)} & \multicolumn{6}{c||}{Kappa ($\kappa$)} &  \multicolumn{6}{c}{Av. Accuracy (AA)}\\\hline
Tile  & \multicolumn{3}{c|}{Pixels} & \multicolumn{3}{c||}{Regions} & \multicolumn{3}{c|}{Pixels} & \multicolumn{3}{c||}{Regions} & \multicolumn{3}{c|}{Pixels} & \multicolumn{3}{c}{Regions} \\
\#    & {CNN} & {CRF}& {2L$\lightning$CRF} & {CNN} & {CRF}& {2L$\lightning$CRF} & {CNN} & {CRF}& {2L$\lightning$CRF} & {CNN} & {CRF}& {2L$\lightning$CRF} &  {CNN} & {CRF}& {2L$\lightning$CRF} &  {CNN}& {CRF} & {2L$\lightning$CRF} \\
\hline
Tile \#4 & 87.3 	& 87.4& 87.7 	& 87.5  & 87.7& 87.7   	& 0.80 &0.80& 0.80&		 0.81& 0.81 	& 0.81& 79.2 & 79.3	& 79.8 & 80.6& 80.7& 80.7  \\
Tile \#6 & 77.8 	& 78.0& 78.9 & 79.3	& 79.6& 80.0 	& 0.68&0.68& 0.69& 		0.70	 &0.70&0.71  & 65.2&65.2&65.4& 65.9&65.9& 66.2  \\
\hline
Avg.$^\dagger$ & 82.6  &82.7& 83.3 \green{(+0.6)}&83.3	&83.5& 83.7	\green{(+0.2)} & 0.77&0.77& 0.78 \green{(+0.01)} & 0.78 &0.79 & 0.79 & 75.2&75.3& 75.6 \green{(+0.3)}& 76.4&76.4& 76.5 \green{(+0.1)} \\
Ov.$^*$ & 82.6 	&82.7& 83.3 \green{(+0.6)}& 83.4	&83.7& 83.8  \green{(+0.1)} & 0.74&0.74&0.75 \green{(+0.01)}& 0.75 &0.76& 0.76 & 72.2&72.2& 72.6 \green{(+0.4)} & 73.3&73.3& 73.4 \green{(+0.1)} \\
\hline
\multicolumn{9}{l}{$^\dagger$ metric over all test samples.}\\
\multicolumn{9}{l}{$^*$ average over the metric  of the 5 test tiles.}\\
\end{tabular}
}
\end{table*}

%

\section{Conclusion}\label{sec:c}

In this paper, we proposed a probabilistic discriminative graphical
model relying on a conditional random fields formulation for the
fusion of land-cover and land-use classification results from very
high resolution remote sensing images. The system is able to find
agreement between probabilistic decisions with multiple spatial
supports. We explored the fusion of pixel- and region-based
{spatial supports} within an energy minimization framework,
where each individual classification result is tributary of i) the
posterior distribution of the land cover classes at the single
instance level (the pixel or the region), ii) the spatial smoothness
of the predictions (i.e. the consistency of the prediction among the
spatial neighbors), and iii) the smoothness across supports (i.e. the
consistency of the predictions between a region and the pixels
composing the region itself). These three goals are addressed jointly
within a conditional random field model with connections {across
  layers corresponding to different spatial support
  representations}. It is also proven that the proposed two-layer
model can be flattened into a single CRF, for which energy
minimization can be addressed efficiently with standard energy
minimization solvers.

Applications to two very high resolution benchmark datasets showed the
potential of the approach that, using common models (we considered
random forests and convolutional neural networks), can improve the
final maps consistently and joining the spatial detail of the pixel
support with the geometrical object accuracy of the region support. In
the future, we would like to test the 2L$\lightning$CRF model {for
  the} fusion of multitemporal data, multiple segmentations ($> 2$),
or for applications involving  different classes to be predicted for
different spatial supports.
{Furthermore, although the impact of the weight parameters of
  the proposed CRF model on the resulting performance was limited, it
  would also be interesting to automatically optimize their values,
  for example using log-likelihood-like (e.g. through
  pseudo-likelihood approximations or the expectation-maximization
  algorithm) or mean-square-error
  concepts~\cite{Alejandro,IoBrunoHK,JosianeZoltan}.}


\section*{Acknowledgments}
This work was supported in part by the Swiss National Science Foundation, via the
  grant 150593 ``Multimodal machine learning for remote sensing information fusion'' (http://p3.snf.ch/project-150593). The authors would also like to thank the Belgian Royal Military Academy, for acquiring and providing the Zeebruges data used in this study, ONERA -- The French Aerospace Lab, for providing the corresponding ground-truth data~\cite{Lag15}, and the IEEE GRSS Image Analysis and Data Fusion Technical Committee.

\vspace{-3mm}

\bibliographystyle{IEEEtran}
\bibliography{hcrf}

\end{document}